\documentclass[11pt,a4paper]{article}
\pdfoutput=1
\usepackage{jheppub}
\usepackage{slashed}
\usepackage{multirow}
\usepackage{color}
\usepackage{mathtools}

\newcommand{\slv}{\raise.15ex\hbox{$/$}\kern-.53em\hbox{$v$}}
\newcommand{\slF}{\raise.15ex\hbox{$/$}\kern-.53em\hbox{$F$}}
\newcommand{\slL}{\raise.15ex\hbox{$/$}\kern-.53em\hbox{$L$}}
\newcommand{\slP}{\raise.15ex\hbox{$/$}\kern-.53em\hbox{$P$}}
\newcommand{\slp}{\raise.15ex\hbox{$/$}\kern-.53em\hbox{$p$}}
\newcommand{\slq}{\raise.15ex\hbox{$/$}\kern-.53em\hbox{$q$}}
\newcommand{\slR}{\raise.15ex\hbox{$/$}\kern-.53em\hbox{$R$}}
\newcommand{\slQ}{\raise.15ex\hbox{$/$}\kern-.53em\hbox{$Q$}}
\newcommand{\slK}{\raise.15ex\hbox{$/$}\kern-.53em\hbox{$K$}}
\newcommand{\slk}{\raise.15ex\hbox{$/$}\kern-.53em\hbox{$k$}}
\newcommand{\slD}{\raise.15ex\hbox{$/$}\kern-.53em\hbox{$D$}}
\newcommand{\slC}{\raise.15ex\hbox{$/$}\kern-.53em\hbox{$C$}}
\newcommand{\slA}{\raise.15ex\hbox{$/$}\kern-.53em\hbox{$A$}}
\newcommand{\slSigma}{\raise.15ex\hbox{$/$}\kern-.53em\hbox{$\Sigma$}}
\newcommand{\slpartial}{\raise.15ex\hbox{$/$}\kern-.53em\hbox{$\partial$}}
\newcommand{\slcalP}{\raise.15ex\hbox{$/$}\kern-.63em\hbox{$\cal P$}}

\newcommand{\beq}{\begin{eqnarray}}
\newcommand{\eeq}{\end{eqnarray}}
\newcommand{\be}{\begin{eqnarray*}}
\newcommand{\ee}{\end{eqnarray*}}

\DeclareMathOperator{\tr}{tr}

\title{KLWMIJ Reggeon Field Theory beyond the large $N_c$ limit. }
\author[a]{Tolga Altinoluk,}
\author[a]{N\'estor Armesto,}
\author[b]{Alex Kovner,}
\author[c,d]{Eugene Levin,}
\author[e]{and Michael Lublinsky}
\affiliation[a]{Departamento de F\'isica de Part\'iculas and IGFAE,
Universidade de Santiago de Compostela,
E-15706 Santiago de Compostela,
Galicia-Spain}
\affiliation[b]{Physics Department, University of Connecticut, 2152 Hillside Road, Storrs, CT 06269, USA}
\affiliation[c]{Departemento de F\'isica, Universidad T\'ecnica Federico Santa Mar\'ia, and Centro Cient\'ifico-\\
Tecnol\'ogico de Valpara\'iso, Avda. Espana 1680, Casilla 110-V, Valpara\'iso, Chile}
\affiliation[d]{Department of Particle Physics, Tel Aviv University, Tel Aviv 69978, Israel}
\affiliation[e]{Physics Department, Ben-Gurion University of the Negev, Beer Sheva 84105, Israel}

\abstract{We extend the analysis of KLWMIJ evolution in terms of QCD Reggeon fields beyond leading order in the $1/N_c$ expansion. We show that there is only one type of corrections to the leading order Hamiltonian discussed in \cite{last}. These are terms linear in original Reggeons and quadratic in conjugate Reggeon operators. All of these have the interpretation as vertices of the "`merging"' type $2\rightarrow 1$, where two Reggeons merge into one. Importantly, the triple Pomeron merging vertex does not emerge from the KLWMIJ Hamiltonian. 
We show that, although in the range of applicability of the KLWMIJ Hamiltonian these merging terms are subleading in $N_c$, in the dense-dense regime they all become of the same (leading) order in $N_c$. In this regime vertices involving higher Reggeons are enhanced by inverse powers of the coupling constant.}

\begin{document}
\maketitle

\pagestyle{empty}
\newpage

\mbox{}

\pagestyle{plain}

\setcounter{page}{1}

\section{Introduction and Conclusion}
\label{sec:intro}
In recent papers \cite{last,last1} we have studied in some detail the relation between the CGC \cite{jimwlk} approach to the high energy evolution of hadronic observables in the guise of JIMWLK \cite{jimwlk} and KLWMIJ \cite{klwmij} equations, and the  Reggeon Field Theory \cite{bartels, braun}. In \cite{last} we have shown that the KLWMIJ Hamiltonian at leading order in the $1/N_c$ expansion can be written as a theory of interacting Reggeons. This theory very simply encodes the Bartels' triple Pomeon vertex, but in addition to Pomerons it contains other Reggeon states, at least some of which play an important role at high energy. In \cite{last1} we have considered a more general Hamiltonian \cite{aklp} which includes the effects of large Pomeron loops, and have shown that it defines a self dual \cite{duality} Reggeon Field Theory. The analysis of \cite{last, last1} was performed in the large $N_c$ limit, although as discussed in \cite{last1} the large $N_c$ counting at high energy is peculiar, as the same  Reggeon field is of different order in  $1/N_c$ in different energy regimes.

In this paper we continue our analysis of the Reggeon Field Theory representation of KLWMIJ evolution, $H_{KLWMIJ}^{RFT}$, this time including all  terms subleading in $N_c$. Our aim is to understand which type of Reggeon interactions are contained in the KLWMIJ Hamiltonian beyond leading order.  
This is important in order to understand whether the theory can be eventually cast into a form of reaction-diffusion process \cite{Munier:2009pc, Kolevatov:2011nf}.
Another reason for the study is to further  clarify the evolution of higher order operators which are 
relevant for correlation-type observables, such as the ridge \cite{ddgjlv, Kovner:2012jm, LR}. Importance of such corrections subeading in $1/N_c$ has been emphasized in \cite{aj}.

Ou results can be summarized in the following way. We show that no $1/N_c^2$ corrections arise to Reggeon propagator terms and to any of the  $1\rightarrow 2$ splitting vertices involving Reggeon operators. The nonvanishing corrections to the RFT Hamiltonian are all of order $1/N_c^2$, and all of them are of the type of $2\rightarrow 1$ merging vertices. 
There is however no $2\rightarrow 1$ triple Pomeron vertex which we normally associate with the JIMWLK Hamiltonian. 

This large $N_c$ counting pertains to the dilute-dense regime which is the appropriate regime for the application of $H_{KLWMIJ}$. It is nevertheless of some interest to rewrite the Hamiltonian in terms of the operators relevant to the dense-dense regime. This involves changing the basis from the Reggeon conjugates to dual Reggeon operators, as suggested in \cite{last1}. When this transformation is performed in $H_{KLWMIJ}^{RFT}$ we find that all vertices (splitting and merging) become of the same, leading order at large $N_c$. We also find that the vertices involving higher Reggeons are enhanced by inverse powers of the 't Hooft coupling constant.

The plan of this paper is the following. In Section 2 we recall the KLWMIJ Hamiltonian, and derive $H_{KLWMIJ}^{RFT}$ including all orders in $1/N_c^2$, in terms of multipole operators and their conjugates. In Section 3 we discuss the definition of $n$-point Reggeon operators in terms of $n$-point multipoles, generalizing the construction of \cite{last}. In Section 4 we discuss the $N_c$ counting in the dense-dense limit. Appendices contain technical details of our derivations.

\section{KLWMIJ Reggeon Field Theory beyond large $N_c$}
\label{sec:evolution}
The KLWMIJ evolution equation is a functional evolution equation for the weight functional $W[\delta/\delta\rho]$. This weight functional $W$ represents the probability density to find a given configuration of charge density in the projectile. The expectation value of any observable ${\cal O}$ is given by 
\beq
\label{avrgdO}
\langle {\cal O}\rangle=\langle \int d\rho \delta(\rho)W[\delta/\delta\rho]{\cal O}[\rho,\alpha]\rangle_{\alpha} \; .
\eeq
Here, $\rho$ is the color charge density of the projectile. We have allowed for the dependence of operator $\cal O$ on the color field of the target $\alpha$, which also  has to be averaged over. The averaging over the target field $\alpha$ is determined solely by the properties of the target, and is not directly relevant to our discussion, since we ascribe all evolution to the projectile wave function.

An example of an interesting observable is the $S$-matrix of a projectile dipole, given by 
\beq
S=\left \langle \int d\rho \delta(\rho) \frac{1}{N_c}\tr[R_x R^{\dagger}_y] e^{i\int_z g^2\rho(z)\alpha(z)}\right \rangle\; .
\eeq
Here $R_x$ is the unitary matrix in the fundamental representation, that represents the scattering amplitude of a quark at transverse point $x$,
\beq
R_x=e^{t^a\frac{\delta}{\delta\rho^a_x}}\; ,
\eeq 
with $t^a$ is the $SU(N_c)$ generator in the fundamental representation.  
A general dilute projectile is not necessarily a single dipole but is rather composed of a small number of color neutral objects. The scattering matrix of such a projectile depends on dipole, quadrupole operators and so on.  These are $SU_L(N_c)\times SU_R(N_c)$ invariant singlets.
Other observables besides the scattering matrix, such as gluon production cross section, gluon correlations, $\dots$, depend on the same color singlets\footnote{This statement is not strictly  precise. In general the gluon production cross section depends on objects which are symmetric only under the diagonal subgroup of the $SU_L(N_c)\times SU_R(N_c)$ symmetry of $H_{KLWMIJ}$ and are not necessarily completely $SU_L(N_c)\times SU_R(N_c)$ symmetric, see for example \cite{cutpom}. Nevertheless, in many cases the evolution of $SU_L(N_c)\times SU_R(N_c)$ symmetric operators is sufficient to also describe the energy dependence of these more general observables \cite{last1}.}. The weight functional $W$ is thus some functional of these color neutral objects:
\beq
W\to W[d_{12}, Q_{1234},...] \; ,
\eeq
where the dipoles and quadrupoles are defined as
\beq
&&d_{12}=\frac{1}{N_c}\tr[R_1R^{\dagger}_2]\; , \nonumber\\
&&Q_{1234}=\frac{1}{N_c}\tr[R_1R^{\dagger}_2R_3R^{\dagger}_4]
\eeq
and similarly for higher level operators.
Here we introduced a shorthand notation for the transverse coordinates $x_1,x_2,...,x_n\to1,2,...,n$.

The evolution of any observable to higher energies in the KLWMIJ approximation is given by 
\beq
\frac{d{\langle {\cal O}\rangle}}{dY}=\int d\rho \delta(\rho)W[\delta/\delta\rho]H_{KLWMIJ}{\cal O}[\rho] \; ,
\eeq
where the explicit form of the KLWMIJ Hamiltonian reads
\beq
H_{KLWMIJ}=\frac{\alpha_s}{2\pi^2}\int_{x,y,z}K_{xyz}\big\{J_L^a(x)J_L^a(y)+J_R^a(x)J_R^a(y)-2J_L^a(x)R_A^{ab}(z)J_R^b(y)\big\} \; .
\eeq 
Here $K_{xyz}$ is the Weizs\"acker-Williams kernel 
\beq
K_{xyz}=\frac{(x-z)_i}{(x-z)^2}\frac{(y-z)_i}{(y-z)^2} \; .
\eeq
As long as we are interested in the action of the KLWMIJ Hamiltonian on gauge invariant quantities (invariant under the $SU_L(N_c)\times SU_R(N_c)$ transformation), the kernel $K_{xyz}$ can be replaced by the dipole kernel 
\beq
K_{xyz}\to-\frac{1}{2}M_{xyz} \: \:  , \: \:\: \:\: \: \: \:\: \: M_{xyz}=\frac{(x-y)^2}{(x-z)^2(y-z)^2} \; .
\eeq
$J_L^a(x)$ and $J_R^a(x)$ are the left and right rotation generators  
\beq
&&J_L^a(x)=\tr\bigg[ \frac{\delta}{\delta R^T_x}t^aR_x\bigg]-\tr\bigg[ \frac{\delta}{\delta R^*_x}R^{\dagger}_xt^a\bigg]\; ,\\
&&J_R^a(x)=\tr\bigg[ \frac{\delta}{\delta R^T_x}R_xt^a\bigg]-\tr\bigg[ \frac{\delta}{\delta R^*_x}t^aR^{\dagger}_x\bigg] \; .
\eeq
The action of the right and left rotation generators on the unitary matrix $R(x)$ are:
\beq
\label{Laction}
&&J_L^a(y)R_x=t^aR_x\delta(y-x) \; ,\hspace{2.5cm}  J_L^a(y)R^{\dagger}_x=-R^{\dagger}_xt^a\delta(y-x) \; .\\
\label{Raction}
&&J_R^a(y)R_x=R_xt^a\delta(y-x) \; , \hspace{2.5cm}  J_R^a(y)R^{\dagger}_x=-t^aR^{\dagger}_x\delta(y-x) \; .
\eeq
Our goal now is to represent the action of $H_{KLWMIJ}$ on a functional which depends on general color singlet operators, similarly to what has been done in \cite{remarks} for $W$ which depends on dipoles only. We start off by assuming for simplicity, that $W$ is a function of $d$ and $Q$ only.

The Hamiltonian is quadratic in the $SU(N)$ rotation generators, and one should simply act with the two rotations sequentially. For simplicity we first consider only a dependence on $d$ and $Q$.
The action of the first rotation operator can be represented as
\beq
\label{1action}
J_{L(R)}^a(y)W[d,Q]=\big[J_{L(R)}^a(y)d_{12}\big]\frac{\delta W}{\delta d_{12}}+\big[J_{L(R)}^a(y)Q_{1234}\big]\frac{\delta W}{\delta Q_{1234}} \; .
\eeq
The integration over the coordinates $1,2$ and $1,2,3,4$ is implicitly assumed in Eq.(\ref{1action}).
Acting with the second generator we obtain, for example
\beq
&&\hspace{-1.2cm}J_R^a(x)J_L^a(y)W[d,Q]=\big[J_R^a(x)J_L^a(y)d_{12}\big]\frac{\delta W}{\delta d_{12}}+\big[J_R^a(x)J_L^a(y)Q_{1234}\big]\frac{\delta W}{\delta Q_{1234}} \nonumber\\
&&\hspace{0.1cm}+\frac{1}{2}\bigg( \big[J_R^a(x)d_{12}\big]\big[J_L^a(y)d_{34}\big]+\big[J_R^a(x)d_{34}\big]\big[J_L^a(y)d_{12}\big]
\bigg)\frac{\delta^2 W}{\delta d_{12}\delta d_{34}}\nonumber\\
&&\hspace{0.1cm}+\frac{1}{2}\bigg( \big[J_L^a(y)d_{12}\big]\big[J_R^a(x)Q_{3456}\big]+\big[J_R^a(x)d_{12}\big]\big[J_L^a(y)Q_{3456}\big]
\bigg)\frac{\delta^2 W}{\delta d_{12}\delta Q_{3456}}\nonumber\\
&&\hspace{0.1cm}+\frac{1}{2}\bigg( \big[J_R^a(x)Q_{1234}\big]\big[J_L^a(y)Q_{5678}\big]+\big[J_R^a(x)Q_{5678}\big]\big[J_L^a(y)Q_{1234}\big]
\bigg)\frac{\delta^2 W}{\delta Q_{1234}\delta Q_{5678}} \; ,
\eeq
where again all coordinates appearing twice are integrated over.

The action of the left and right rotation operators on dipoles and quadrupoles is straightforwardly calculated with the help of the Eqs. \eqref{Laction} and \eqref{Raction}. To simplify the resulting expressions we use the completeness relation 
\beq
t^a_{\alpha \beta} t^a_{\gamma\lambda}=\frac{1}{2}\bigg( \delta_{\alpha\lambda} \delta_{\beta\gamma}-\frac{1}{N_c}\delta_{\alpha\beta} \delta_{\gamma\lambda}\bigg)
\eeq
and the identity
\beq
R^{ab}_z=2\tr[t^aR_zt^bR^{\dagger}_z] \; .
\eeq
After some algebra we find that the action of the KLWMIJ Hamiltonian on $W[d,Q]$ can be represented as the sum of the following terms :
\beq
\label{decomposedH}
H_{KLWMIJ}=H_d+H_Q+\frac{1}{N_c^2}\big( H_{dd}+H_{dQ}+H_{QQ}\big) \; .
\eeq
The leading $N_c$ terms reproduce and generalize the Mueller's dipole model \cite{Mueller} and have been derived in \cite{remarks, last}:
\beq
\label{Hd}
&&\hspace{-1.2cm}H_d=-\frac{\bar{\alpha_s}}{2\pi}\int_{z,12}M_{12z}\big( d_{1z}d_{z2}-d_{12}\big)\frac{\delta}{\delta d_{12}}\; ,\\
\label{HQ}
&&\hspace{-1.2cm}H_Q=-\frac{\bar{\alpha_s}}{2\pi}\int_{z,1234}\bigg\{-\big(M_{12z}+M_{34z}-L_{1342z} \big)Q_{1234}-L_{1432z}d_{12}d_{34}-L_{1234z}d_{14}d_{32}\\
&&+L_{1214z}d_{1z}Q_{z234}+L_{1232z}d_{z2}Q_{1z34}+L_{3234z}d_{3z}Q_{12z4}+L_{1434z}d_{z4}Q_{123z}\bigg\}\frac{\delta}{\delta Q_{1234}}\; ,\nonumber
\eeq
where $\bar{\alpha}_s\equiv \alpha_sN_c/\pi$ and 
\beq
L_{xyuvz}&=&\bigg[ \frac{(x-z)_i}{(x-z)^2}-\frac{(y-z)_i}{(y-z)^2}\bigg]\bigg[ \frac{(u-z)_i}{(u-z)^2}-\frac{(v-z)_i}{(v-z)^2}\bigg]\nonumber\\
&=&\frac{1}{2}\big[M_{yuz}+M_{xvz}-M_{yvz}-M_{xuz} \big] \; .
\eeq
The terms subleading in $N_c$ account for color correlations and are given by 
\beq\label{hdd}
&&\hspace{-1.5cm}H_{dd}=-\frac{{\bar\alpha_s}}{2\pi}\int_{z1234}L_{1243z}\frac{1}{2}\bigg\{\big[Q_{1234}-X_{12z43z}\big]+\big[Q_{1234}-X_{12z43z}\big]^T\bigg\}\frac{\delta^2}{\delta d_{12}\delta d_{34}} ,\\
&&\hspace{-1.5cm}H_{dQ}=-\frac{{\bar\alpha_s}}{2\pi}\int_{z123456}L_{1263z}\bigg\{\big[X_{123456}-\Sigma_{3456z21z}\big]\frac{\delta^2}{\delta d_{12}\delta Q_{3456}}\nonumber\label{hdq}\\
&&\hspace{5.5cm}+\big[X_{123456}-\Sigma_{3456z21z}\big]^T\frac{\delta^2}{\delta d_{12}\delta (Q_{3456})^T}\bigg\} ,\\
&&\hspace{-1.5cm}H_{QQ}=-\frac{{\bar\alpha_s}}{2\pi}\int_{z12345678}2L_{1485z}\bigg\{\big[\Sigma_{12345678}-Q_{1234}Q_{5678}\big]\frac{\delta^2}{\delta Q_{1234}\delta Q_{5678}}\nonumber\\
&&\hspace{3.6cm}-\big[\Omega_{1234z8765z}-Q_{1234}(Q_{5678})^T\big]\frac{\delta^2}{\delta Q_{1234}\delta (Q_{5678})^T}\nonumber\\
&&\hspace{3.6cm}+\big[\Sigma_{12345678}-Q_{1234}Q_{5678}\big]^T\frac{\delta^2}{\delta (Q_{1234})^T\delta (Q_{5678})^T}\nonumber\\
&&\hspace{3.6cm}-\big[(\Omega_{1234z8765z})^T-(Q_{1234})^TQ_{5678}\big]\frac{\delta^2}{\delta (Q_{1234})^T\delta Q_{5678}}\bigg\}\label{hqq} ,
\eeq
where we defined the higher multiplets as
\beq
X_{123456}&=&\frac{1}{N_c}\tr[R_1R^{\dagger}_2R_3R^{\dagger}_4R_5R^{\dagger}_6],\\
\Sigma_{12345678}&=&\frac{1}{N_c}\tr[R_1R^{\dagger}_2R_3R^{\dagger}_4R_5R^{\dagger}_6R_7R^{\dagger}_8],\\
\Omega_{12345\bar{1}\bar{2}\bar{3}\bar{4}\bar{5}}&=&\frac{1}{N_c}\tr[R_1R^{\dagger}_2R_3R^{\dagger}_4R_5R^{\dagger}_{\bar{1}}R_{\bar{2}}R^{\dagger}_{\bar{3}}R_{\bar{4}}R^{\dagger}_{\bar{5}}],
\eeq
and have introduced the notation $Q^T_{1234}\equiv Q_{4321}$.

Two important properties stand out in the equations above. There are no $1/N_c^2$ corrections to the term linear in functional derivatives (conjugate momenta) in Eq.(\ref{decomposedH}). The only correction term is quadratic in functional derivatives, and it itself also has no further $1/N_c$ corrections. We now show that these properties remain true even if one includes higher Reggeons in the evolution.

The generalization to higher Reggeons is straightforward, albeit lengthy. We give details of the derivation in Appendix 1, and here only present the final result.
Consider a general probability density distribution, 
\beq
W=W[\{O_{2n}\}], \ \ \ \ \ \ \ O_{2n}= \frac{1}{N_c}\tr[R_1R^{\dagger}_2...R_{2n-1}R^{\dagger}_{2n}]\; .
\eeq
We show in Appendix 1 that the action of $H_{KLWMIJ}$ on such a functional can be written as
\beq
\label{genralized_decomposedH}
H_{KLWMIJ}=\sum_nH_{O_{2n}}+\frac{1}{N_c^2}\sum_{n,m}H_{O_{2n}O_{2m}}
\eeq
where the two terms are linear and quadratic in functional derivatives respectively :
\beq\label{ho2n}
&&H_{O_{2n}}=\frac{\bar{\alpha_s}}{2\pi}\int_{z,1,\ldots,2n} \frac{\delta}{\delta O_{2n}(1,\ldots,2n)}\Bigg\{\ \ \sum_{j=1}^{n}\frac{1}{2}\left[M_{2j-1,2j,z}+M_{2j-1,2j-2,z}\right]
O_{2n}(1,\ldots,2n)\nonumber\\
&&+\sum_{j=2}^{n}\sum_{k=1}^{j-1}L_{2j-1,2j-2,2k-1,2k-2;z} O_{2(k+n-j)}(1,\ldots,2k-2,2j-1,\ldots,2n)\nonumber\\
&&\hspace{8cm}\times O_{2(j-k)}(2k-1,\ldots,2j-2)
\nonumber\\
&&+\sum_{j=2}^{n}\sum_{k=1}^{j-1}L_{2j-1,2j,2k-1,2k;z} O_{2(k+n-j)}(1,\ldots,2k-1,2j,\ldots,2n)\nonumber\\
&&\hspace{8cm}\times O^T_{2(j-k)}(2k,\ldots,2j-1)
\nonumber\\
&&-\sum_{j=1}^{n}\sum_{k=1}^{j}L_{2j-1,2j,2k-1,2k-2;z} O_{2(k+n-j)}(1,\ldots,2k-2,z,2j,\ldots,2n)\nonumber\\
&&\hspace{8cm}\times O_{2(j-k+1)}(2k-1,\ldots,2j-1,z)
\nonumber\\
&&-\sum_{j=1}^{n}\sum_{k=j+1}^{n}L_{2j-1,2j,2k-1,2k-2;z} O_{2(j+n-k+1)}(1,\ldots,2j-1,z,2k-1,\ldots,2n)\nonumber\\
&&\hspace{8cm}\times O_{2(k-j)}(z,2j,\ldots,2k-2)\Bigg\} \; ,
\eeq


\beq\label{ho2ho2m}
&&H_{O_{2n}O_{2m}}=\frac{\bar{\alpha_s}}{4\pi}\int_{z,1\ldots,2n,\bar 1,\ldots,2\bar m} \frac{\delta^2}{\delta O_{2n}(1,\ldots,2n)\delta O_{2\bar{m}}(\bar 1,\ldots,2\bar m)}
\sum_{j=1}^{n}\sum_{k=\bar{1}}^{\bar{m}}\\
&&\times\Bigg\{L_{2j-1,2j-2,2k-\bar{1},2k-\bar{2};z}
O_{2(n+\bar{m})}(1, \ldots , 2j-2, 2k-\bar{1} , \ldots , 2\bar{m}, \bar{1} , \ldots , 2k-\bar{2} , 2j-1 , \ldots , 2n) \nonumber\\
&&+L_{2j-1,2j,2k-\bar{1},2k-\bar{2};z}O_{2(n+\bar{m})}(1, \ldots, 2j-1, 2k,\ldots, 2\bar{m},\bar{1},\ldots, 2k-\bar{1},2j,\ldots,2n)\nonumber\\
&&-L_{2j-1,2j-2,2k-\bar{1},2k;z}O_{2(n+\bar{m}+1)}(1,\ldots,2j-2,z,2k,\ldots,2\bar{m},\bar{1},\ldots,2k-\bar{1},z,2j-1,\ldots,2n) \nonumber\\
&&-L_{2j-1,2j,2k-\bar{1},2k-\bar{2};z} O_{2(n+\bar{m}+1)}(1,l\dots,2j-1,z,2k-\bar{1},\ldots,2\bar{m},\bar{1},\ldots,2k-\bar{2},z,2j,\ldots,2n) \Bigg\},\nonumber
\eeq
where $y_0\equiv y_{2n}$. Eq.\eqref{ho2n} generalizes the evolution of a single  dipole (or single quadrupole) to arbitrary multipoles. In Appendix 2, we present the explicit expressions for the evolution of $6$-point function ($X$) and $8$-point function ($\Sigma$).

%

\section{Construction of Reggeons }
\label{sec:reggeons}
In principle, Eqs.(\ref{ho2n}) and (\ref{ho2ho2m}) give the complete representation of $H_{KLWMIJ}$ acting on $SU_L(N_c)\times SU_R(N_c)$ invariant operators. 
It is however useful to define Reggeon fields as irreducible representations of the discrete symmetries of the problem. This was done for the dipole and quadrupole in \cite{last}. In this section we extend this discussion to include higher Reggeons.
 
 As explained in detail in \cite{last} and \cite{reggeon}, in addition to the $SU_L(N_c)\times SU_R(N_c)$ continuous symmetry group, $H_{KLWMIJ}$ also posesses the discrete signature $Z_2$ symmetry, $R\to R^{\dagger}$, and the discrete charge conjugation symmetry, $R\to R^*$. In \cite{last} the Reggeon fields were constructed as singlets under $SU_L(N_c)\times SU_R(N_c)$ and either even or odd under signature and charge conjugation symmetries. 
In addition, in \cite{last} we have constructed the $Q^{++}$-Reggeon\footnote{The $Q^{++}$-Reggeon was called $B$-Reggeon in \cite{last} and \cite{last1}.} using a simple subtraction such that the evolution equation for the $Q^{++}$-Reggeon does not contain a single Pomeron source term. We generalize this subtraction procedure here.

The decomposition for the two point function is straightforward \cite{last}. The Pomeron and the Odderon fields are defined as
\beq
\label{P}
P_{12}=\frac{1}{2}\big[2-d_{12}-d_{21}\big],
\eeq
\beq
\label{O}
O_{12}=\frac{1}{2}\big[d_{12}-d_{21}\big].
\eeq
The Pomeron is even and the Odderon is odd under both signature and charge conjugation. 
The evolution equations for $P$ and $O$ are
\beq
\label{Pev}
\frac{d}{dY}P_{12}=\frac{\bar{\alpha}_s}{2\pi}\int_zM_{12z}\big[P_{1z}+P_{z2}-P_{12}-P_{1z}P_{z2}-O_{1z}O_{z2}\big],
\eeq
\beq
\label{Oev}
\frac{d}{dY}O_{12}=\frac{\bar{\alpha}_s}{2\pi}\int_zM_{12z}\big[O_{1z}+O_{z2}-O_{12}-O_{1z}P_{z2}-P_{1z}O_{z2}\big].
\eeq
Now let us consider the quadrupole. Decomposing it into eigenstates of discrete symmetries, we define $Q^{++}$, $Q^{+-}$, $Q^{--}$ and $Q^{-+}$ Reggeons :
\beq
\label{4BR}
\hspace{-0.5cm}Q^{++}_{1234}=\frac{1}{4}\big[4\!-\!Q_{1234}\!-\!Q_{4123}\!-\!Q_{3214}\!-\!Q_{2143}\big]-\big[P_{12}-P_{13}+P_{14}+P_{23}-P_{24}+P_{34} \big] ,
\eeq
\beq
\label{4CR}
Q^{+-}_{1234}=\frac{1}{4}\big[Q_{1234}+Q_{4123}-Q_{3214}-Q_{2143}\big],
\eeq
\beq
\label{4DM}
Q^{--}_{1234}=\frac{1}{4}\big[Q_{1234}-Q_{4123}-Q_{3214}+Q_{2143}\big],
\eeq
\beq
\label{4DP}
Q^{-+}_{1234}=\frac{1}{4}\big[Q_{1234}-Q_{4123}+Q_{3214}-Q_{2143}\big].
\eeq
Three independent $Q^{++}$-Reggeon and three independent $Q^{+-}$-Reggeon are 
\beq
Q^{++}_{1234},Q^{++}_{1243}, Q^{++}_{1342} \hspace{1cm}\text{and}\hspace{1cm} Q^{+-}_{1234},Q^{+-}_{1243}, Q^{+-}_{1342}  .
\eeq
The $++$-Reggeon has quantum numbers of the Pomeron. It is signature and charge conjugation even. $+-$-Reggeon is signature even and charge conjugation odd, $-+$ signature odd, charge conjugation even and $--$ has the same quantum numbers as the Odderon - odd under both signature and charge conjugation.
The subtraction of the Pomeron terms in $Q^{++}$ achieves simultaneously two goals. First, when expanded in powers of $\delta/\delta\rho$, the field $Q^{++}$ starts with the term $(\delta/\delta\rho)^4$ as opposed to $Q$ whose expansion starts with $(\delta/\delta\rho)^2$. Second, the evolution equation for $Q^{++}$ does not contain a term linear in $P$, which is in principle allowed due to the identical quantum numbers of $P$ and $Q^{++}$\cite{last}. This evolution equation for completeness is presented in Appendix 2.

Clearly, the $+-$ and $-+$ Reggeons do not require any subtractions, since they do not mix with either the Pomeron or the Odderon. The $--$ Reggeon on the other hand does mix with the Odderon, and its evolution does contain a source due to the Odderon. However, it turns out to be impossible to find a local subtraction of the type of Eq.(\ref{4BR}) which eliminates this source term. In the following we will therefore only consider subtractions in the Pomeron channel.

Moving on to the  six point function, we define the $X^{++}$ Reggeon in the analogous way
\beq
\label{XB}
&&\hspace{-0.8cm}X^{++}_{123456}=\frac{1}{4}\big[4-X_{123456}-X_{612345}-X_{654321}-X_{543216}\big]\nonumber\\
&&-\big[P_{12}\!-\!P_{13}\!+\!P_{14}\!-\!P_{15}\!+\!P_{16}\!+\!P_{23}\!-\!P_{24}\!+\!P_{25}\!-\!P_{26}\!+\!P_{34}\!-\!P_{35}\!+\!P_{36}\!+\!P_{45}\!-\!P_{46}\!+\!P_{56} \big]\nonumber\\
&&-\big[Q^{++}_{1234}\!-\!Q^{++}_{1235}\!+\!Q^{++}_{1236}\!+\!Q^{++}_{1245}\!-\!Q^{++}_{1246}\!+\!Q^{++}_{1256}\!-\!Q^{++}_{1345}\nonumber\\
&&\hspace{2.8cm}\!+\!Q^{++}_{1346}\!-\!Q^{++}_{1356}\!+\!Q^{++}_{1456}\!+\!Q^{++}_{2345}\!-\!Q^{++}_{2346}\!+\!Q^{++}_{2356}\!-\!Q^{++}_{2456}\!+\!Q^{++}_{3456}\big].
\eeq
As shown in  Appendix 2, this particular subtraction achieves the required effect, namely it eliminates source terms proportional to $P$ and $Q^{++}$ from the evolution equation for $X^{++}$. This equation can be found in Appendix 2.
The following ten $X^{++}$ Reggeons associated with given six points, are independent:
\beq
\hspace{-0.7cm}X^{++}_{123456},X^{++}_{123465},X^{++}_{123546},X^{++}_{124356},X^{++}_{124365},X^{++}_{125364},X^{++}_{132456},X^{++}_{132465},X^{++}_{132546},X^{++}_{142536}.
\eeq
Analogously we define
\beq
\label{XC}
&&\hspace{-0.8cm}X^{+-}_{123456}=\frac{1}{4}\big[X_{123456}+X_{612345}-X_{654321}-X_{543216}\big]\nonumber\\
&&-\big[Q^{+-}_{1234}\!-\!Q^{+-}_{1235}\!+\!Q^{+-}_{1236}\!+\!Q^{+-}_{1245}\!-\!Q^{+-}_{1246}\!+\!Q^{+-}_{1256}\!-\!Q^{+-}_{1345}\nonumber\\
&&\hspace{2.6cm}\!+\!Q^{+-}_{1346}\!-\!Q^{+-}_{1356}\!+\!Q^{+-}_{1456}\!+\!Q^{+-}_{2345}\!-\!Q^{+-}_{2346}\!+\!Q^{+-}_{2356}\!-\!Q^{+-}_{2456}\!+\!Q^{+-}_{3456}\big] ,
\eeq
and  ten independent $X^{+-}$-Reggeons are
\beq
\hspace{-0.7cm}X^{+-}_{123456},X^{+-}_{123465},X^{+-}_{123546},X^{+-}_{124356},X^{+-}_{124365},X^{+-}_{125364},X^{+-}_{132456},X^{+-}_{132465},X^{+-}_{132546},X^{+-}_{142536} .
\eeq
The signature odd 6-point Reggeons are given as
\beq\label{XDP}
&&\hspace{-0.8cm}X^{-+}_{123456}=\frac{1}{4}\big[X_{123456}-X_{612345}+X_{654321}-X_{543216}\big]\nonumber\\
&&-\big[Q^{-+}_{1234}\!-\!Q^{-+}_{1235}\!+\!Q^{-+}_{1236}\!+\!Q^{-+}_{1245}\!-\!Q^{-+}_{1246}\!+\!Q^{-+}_{1256}\!-\!Q^{-+}_{1345}\nonumber\\
&&\hspace{2.8cm}\!+\!Q^{-+}_{1346}\!-\!Q^{-+}_{1356}\!+\!Q^{-+}_{1456}\!+\!Q^{-+}_{2345}\!-\!Q^{-+}_{2346}\!+\!Q^{-+}_{2356}\!-\!Q^{-+}_{2456}\!+\!Q^{+-}_{3456}\big] 
\eeq
and finally
\beq
\label{XDM}
X^{--}_{123456}=\frac{1}{4}\big[X_{123456}-X_{612345}-X_{654321}+X_{543216}\big] .
\eeq
Note that we have included subtractions in the $+-$ and $-+$ channels. These eliminate sources proportional to $Q^{+-}$ in the evolution of $X^{+-}$ and similarly for $X^{-+}$.
We do not bother with subtractions in the $--$ channel, since the Odderon source terms cannot be eliminated already in the evolution of $Q^{--}$.

The subtraction structure can be generalized to an arbitrary $n$-point Reggeon by inspection.
Let us first define the unsubtracted  signature and charge conjugation even multipole combinations:
\beq
d^{SS}_{12}=\frac{1}{2}\Big[d_{12}+d_{21}\Big] ,
\eeq
\beq
Q^{SS}_{1234}=\frac{1}{4}\Big[Q_{1234}+Q_{4123}+Q_{3214}+Q_{2143}\Big] ,
\eeq
\beq
X^{SS}_{123456}=\frac{1}{4}\Big[X_{123456}+X_{612345}+X_{654321}+X_{543216}\Big]
\eeq
and, for an arbitrary $2n$ point function,
\beq
&&\hspace{-1cm}O^{SS}_{2n}(1,\ldots,2n)=\frac{1}{4}\Big[O_{2n}(1,\ldots,2n)+O_{2n}(2n,1,\ldots,2n-1)\nonumber\\
&&\hspace{5cm}+O^T_{2n}(1,\ldots,2n)+O^T_{2n}(2n,1,\ldots,2n-1)\Big] .
\eeq
The generalisation of Eqs.\eqref{4BR} and \eqref{XB} for arbitrary $n$ can then be written as
\begin{equation}\label{q+}
O^{SS}_{2n}(1,\ldots,2n)=
1-\Bigg[\sum_{l=0}^{n-1}\sum_{\{i_1<i_2<\ldots<i_{2(n-l)}\}}(-1)^{n-l+\sum_{k=1}^{2(n-l)} i_k}O^{++}_{2(n-l)}(i_1,i_2,\ldots, i_{2(n-l)})\Bigg] .
\end{equation}
Analogously, defining
\begin{eqnarray}
&&\hspace{-2cm}O^{SA}_{2n}(1,\ldots,2n)=\frac{1}{4}\Big[O_{2n}(1,\ldots,2n)+O_{2n}(2n,1,\ldots,2n-1)\nonumber\\
&&\hspace{4cm}-O^T_{2n}(1,\ldots,2n)-O^T_{2n}(2n,1,\ldots,2n-1)\Big],\nonumber\\
&&\hspace{-2cm}O^{AS}_{2n}(1,\ldots,2n)=\frac{1}{4}\Big[O_{2n}(1,\ldots,2n)-O_{2n}(2n,1,\ldots,2n-1)\nonumber\\
&&\hspace{4cm}+O^T_{2n}(1,\ldots,2n)-O^T_{2n}(2n,1,\ldots,2n-1)\Big],\nonumber\\
&&\hspace{-2cm}O^{SA}_2=O^{AS}_2\equiv 0,
\end{eqnarray}
we have
\begin{eqnarray}\label{q-}
&&\hspace{-1cm}O^{SA}_{2n}(1,\ldots,2n)=
1-\Bigg[\sum_{l=0}^{n-1}\sum_{\{i_1<i_2<\ldots<i_{2(n-l)}\}}\hspace{-1cm}(-1)^{n-l+\sum_{k=1}^{2(n-l)} i_k}O^{+-}_{2(n-l)}(i_1,i_2,\ldots, i_{2(n-l)})\Bigg],\nonumber\\
&&\hspace{-1cm}O^{AS}_{2n}(1,\ldots,2n)=
1-\Bigg[\sum_{l=0}^{n-1}\sum_{\{i_1<i_2<\ldots<i_{2(n-l)}\}}\hspace{-1cm}(-1)^{n-l+\sum_{k=1}^{2(n-l)} i_k}O^{-+}_{2(n-l)}(i_1,i_2,\ldots, i_{2(n-l)})\Bigg].
\end{eqnarray}

Expressing Reggeons in terms of multipoles requires some extra algebra.
For $n=1,2,3$ this is straightforward, and we obtain
\beq
P_{12} = 1 - d^{SS}_{12} ,
\eeq
\beq
Q^{++}_{1234} = - 1+ \Big[ d^{SS}_{12}-d^{SS}_{13}+d^{SS}_{14}+d^{SS}_{23}-d^{SS}_{24}+d^{SS}_{34} \Big] -Q^{SS}_{1234} ,
\eeq 
\beq
&&\hspace{-0.5cm}X^{++}_{123456} = 1 - \big[d^{SS}_{12}\!-\!d^{SS}_{13}\!+\!d^{SS}_{14}\!-\!d^{SS}_{15}\!+\!d^{SS}_{16}\!+\!d^{SS}_{23}\!-\!d^{SS}_{24}\!+\!d^{SS}_{25}\!-\!d^{SS}_{26}\!+\!d^{SS}_{34}\!-\!d^{SS}_{35}\nonumber\\
&&\hspace{10cm}\!+\!d^{SS}_{36}\!+\!d^{SS}_{45}\!-\!d^{SS}_{46}\!+\!d^{SS}_{56} \big]\nonumber\\
&&\hspace{1.6cm}
+\big[Q^{SS}_{1234}\!-\!Q^{SS}_{1235}\!+\!Q^{SS}_{1236}\!+\!Q^{SS}_{1245}\!-\!Q^{SS}_{1246}\!+\!Q^{SS}_{1256}\!-\!Q^{SS}_{1345}\!+\!Q^{SS}_{1346}\!-\!Q^{SS}_{1356}\!+\!Q^{SS}_{1456}\nonumber\\
&&\hspace{4.1cm}\!+\!Q^{SS}_{2345}\!-\!Q^{SS}_{2346}\!+\!Q^{SS}_{2356}\!-\!Q^{SS}_{2456}\!+\!Q^{SS}_{3456}\big]-X^{SS}_{123456} .
\eeq
It is obvious that these expressions lend themselves to a generalization for arbitrary $n$ :

\begin{equation}\label{qb}
O^{++}_{2n}(1,\ldots,2n)=
(-1)^{n+1}\Bigg[\sum_{l=0}^{n-1}\sum_{\{i_1<i_2<\ldots<i_{2(n-l)}\}}\hspace{-1cm}(-1)^{\sum_{k=1}^{2(n-l)} i_k}O^{SS}_{2(n-l)}(i_1,i_2,\ldots, i_{2(n-l)})+1\Bigg] .
\end{equation}
\section{Merging Vertices from the KLWMIJ Hamiltonian}
Using Eqs.(\ref{q+}) and (\ref{qb}) it is straigtforward to rewrite the Hamiltonian Eqs.\eqref{ho2n} and \eqref{ho2ho2m} in terms of the Reggeon fields.
One expresses the multipole operators in terms of the Reggeons using Eq.(\ref{q+}) and the functional derivatives via
\beq\label{sub}
&&\hspace{-2cm}\frac{\delta}{\delta O_{2n}(1,\ldots,2n)}=\sum_{l=0}^{\infty}\int_{2n+1,\ldots,2(n+l)}\sum_{\{i_1,\ldots,i_{2(n+l)}\}}\sum_{\alpha,\beta=\pm}\nonumber\\
&&\hspace{2cm}\times\frac{\delta O^{\alpha\beta}_{2(n+l)}(i_1,\ldots,i_{2(n+l)})}{\delta O_{2n}(1,\ldots,2n)}\frac{\delta}{\delta  O^{\alpha\beta}_{2(n+l)}(i_1,\ldots,i_{2(n+l)})} .
\eeq
We are not going to perform the substitution Eq.(\ref{sub}) explicitly. We only note that it does not change the nature of the terms in the Hamiltonian. The one important result of this substitution is that it eliminates the terms of the form
\begin{equation}
O^{++}_{2(n-l)}\frac{\delta}{\delta O^{++}_{2n}}, \ \ O^{+-}_{2(n-l)}\frac{\delta}{\delta O^{+-}_{2n}}, \ \ O^{-+}_{2(n-l)}\frac{\delta}{\delta O^{-+}_{2n}}
\end{equation}
which otherwise appear in Eq.(\ref{ho2n}) when one of the multipole operators is taken to be unity. It thus leads to partial diagonalization of the equations, in the sense that the evolution of a level $2n$ Reggeon does not contain a term linear in level $2(n-k)$ Reggeons.

The structure of the Hamiltonian thus can be described as follows. The leading $N_c$ terms are of two types. First, there are homogeneous terms of the type
\begin{equation}
 O^{++}_{2n}\frac{\delta}{\delta O^{++}_{2n}}, \ \ \ \ etc.
\end{equation}
For $++$, $+-$ and $-+$ Reggeons these are diagonal, i.e. they mix only Reggeons on the same level $n$. 

The second type of terms are $1\rightarrow 2$ splitting vertices, contained in Eq.(\ref{ho2n}). These are akin to the $1\rightarrow 2$ Pomeron splitting vertices and generalize it to higher Reggeons. The most interesting such vertex corresponds to splitting of the $Q^{++}$ Reggeon into two Pomerons, and contributes to the effective four Pomeron vertex as discussed in \cite{last1}.

Finally the third type of terms is of order $1/N_c^2$, and is of a different nature. These are ``merging'' vertices of the $2\rightarrow 1$ type. Note that there is no vertex corresponding to transition from two Pomeron to one Pomeron.  This is straightforward to see examining Eqs.\eqref{hdd}, \eqref{hdq}, \eqref{hqq}. In Eq.\eqref{hdd} all the single Pomeron terms cancel in the factor that multiply the functional derivatives, while Eqs. \eqref{hdq} and \eqref{hqq} do not contain a term with two functional derivatives with respect to the Pomeron. This is natural, since we expect to find this vertex in the JIMWLK Hamiltonian which describes the dense projectile regime, and not in the KLWMIJ Hamiltonian. On the other hand Eqs.\eqref{hdd}, \eqref{hdq}, \eqref{hqq} do give rise, for example  to a vertex of the type $PP\rightarrow Q^{++}$.

The order in $1/N_c$ assigned to the vertices is based on large $N_c$ counting for the dilute projectile, dense target regime. In this regime all the Reggeons are close to saturation, and thus are of order one. The Reggeon conjugate operators $\delta/\delta O_{2n}$ are also assigned $O(1)$ at large $N_c$. However, as discussed in \cite{last1} the large $N_c$ limit is tricky, and the order in $N_c$ of a given operator depends on whether the target or the projectile is saturated, or both. 
For example, for a dense projectile a dipole scattering on it scatters with probability of order one. This means that the following dual Pomeron operator is of order one in the large $N_c$ limit :
\begin{equation}\label{dualP}
\bar P(x,y)=1-\frac{1}{2N_c}{\rm Tr}[ S^{F\,\dagger} (x)S^F(y)+S^{F\,\dagger}(y)S^F(x)] ,
\end{equation} 
where
\begin{equation}
S^F(x)=e^{ig^2t^a\alpha_P^a(x)}
\end{equation}
and $\alpha_P$ is the color field created by the color charges of the projectile. The dual Pomeron operator is a nonlinear function of the projectile color sources. However, for weak fields it is related to the Pomeron conjugate operator by a simple linear relation
\begin{equation}
\frac{\delta}{\delta P(x,y)}=P^\dagger (x,y)= \frac{4}{g^4}\nabla^2_x\nabla^2_y\bar P(x,y)=\frac{N_c^2}{4\pi^4 \bar{\alpha}_s^2}\nabla^2_x\nabla^2_y\bar P(x,y)
\end{equation}
In the dense projectile limit $\bar P$ remains finite even at large $N_c$. Thus, if this relation can be taken as a guide of the correct $N_c$ behaviour, in the limit of dense projectile, where $\bar P\sim O(1)$, one has $\frac{\delta}{\delta P(x,y)}\sim O(N_c^2)$. Similarly, we have for the 4 point Reggeon \cite{last1}
\begin{eqnarray}\label{bcbardagger}
Q^{++\dagger}(x,y,u,v)&=&\frac{N_c^2}{64\pi^8 \bar{\alpha}_s^4}\nabla^2_x\nabla^2_y\nabla^2_u\nabla^2_v\bar Q^{++}(x,y,u,v),\nonumber \\
Q^{+-\dagger}(x,y,u,v)&=&\frac{N_c^2}{64\pi^8 \bar{\alpha}_s^4}\nabla^2_x\nabla^2_y\nabla^2_u\nabla^2_v\bar Q^{+-}(x,y,u,v).
\end{eqnarray}
The dual Reggeons with the nonvacuum quantum numbers are not important in the projectile saturation regime, since the limiting saturated value of $Q^{+-}$ etc. is zero. However the $++$ conjugate Reggeon is again $O(N_c^2)$. The same is true for higher Reggeons. In general for a $n$-point  $++$ Reggeon we have
\begin{equation}\label{scaling}
O^{++\dagger}_{2n}(1,\ldots,2n)=\frac{4N_c^2}{(4\pi^2)^{2n} \bar{\alpha}_s^{2n}}\nabla^2_1\ldots\nabla^2_{2n}\bar O^{++}_{2n}(1,\ldots2n) .
\end{equation}
Substituting this into the expression for the Hamiltonian, we see that in the regime where both the projectile and the target are dense, all terms in the Hamiltonian are of the order $N_c^2$. 

Another interesting feature of this regime is that the vertices involving higher Reggeons are in fact enhanced by inverse powers of the $\bar{\alpha}_s$, when 't Hooft coupling is small. The vertices scale as\footnote{Note that when the target and the projectile are saturated, $O_{2n}=O(1)$ for any $n$. This follows from their definition in terms of the appropriate multipoles, as all multipoles vanish at saturation.}
\begin{equation} 
V_{splitting}^{n; m,k}\bar O_{2n}O_{2m} O_{2k}\propto \frac{N_c^2}{\bar{\alpha}_s^{2n-1}}, \ \ \ \ \ \ \ V_{merging}^{n; m,k} O_{2n}\bar O_{2m} \bar O_{2k}\propto \frac{N_c^2}{\bar{\alpha}_s^{2m+2k-1}}.
\end{equation}

It is possible that Eq.(\ref{scaling}) is not correct parametrically in the dense limit, since strictly speaking it was deduced in the dilute regime. However if it holds, this suggests an interesting albeit complicated picture. At weak 't Hooft coupling, the higher Reggeons which are less important in the dilute-dilute and dense-dilute regime (since their value is small at small $\bar\alpha_s$) take over the quantum evolution in the dense-dense regime. To make sense of this regime one would then have to somehow resum the contributions of higher Reggeons. In other words one would have to deal with an essentially nonlinear regime of the Reggeon field theory, a level higher than nonlinear QCD which leads to the Reggeon field theory in the first place.

\section{Appendix 1: The derivation of the KLWMIJ Reggeon theory}
\label{sec:app1}
 In this Appendix, we  derive the action of $H_{KLWMIJ}$ on a generic functional of color singlet operators of the type
\beq
O_{2n}= \frac{1}{N_c}\tr[R_1R^{\dagger}_2...R_{2n-1}R^{\dagger}_{2n}].\hspace{1cm}
\eeq
The simplest such operator is the dipole.
The dipole evolution operator $H_d$ given in Eq.\eqref{Hd}, picks up a dipole from the weight functional $W$ and evolves that dipole via the KLWMIJ Hamiltonian \emph{i.e.} $H_d$ can equivalently be written as 
\beq
H_d=\int_{12}\Bigg[H_{KLWMIJ}d_{12}\Bigg]\frac{\delta}{\delta d_{12}} .
\eeq
In the KLWMIJ evolution of a single dipole $d_{12}$ the subleading piece in $1/N_c$ cancels between the real and virtual terms in the Hamiltonian and the exact KLWMIJ evolution of a dipole is all leading $N_c$. The same is true for the evolution of a quadrupole. The color correlators in Eq.\eqref{decomposedH} that are subleading in $N_c$ arise from the evolution of a product of two color neutral objects (two dipoles, one dipole and one quadruple and two quadrupoles). In this case, one of the two color rotation operators acts on one object, and the other acts on the second object generating the order  $1/N_c^2$ term. The terms of order $1/N_c^3$, which a priori could be present, cancel in the final result between the real and virtual contributions of the KLWMIJ Hamiltonian. Hence,  the expression given in Eq.\eqref{decomposedH} is the result of  exact color algebra. In order to generalise this expression one should show that a similar cancellation happens for the generic color neutral object $O_{2n}$.

\subsection{KLWMIJ Evolution of a generic single trace operator}
First, we derive the evolution equation of a single trace operator with arbitrary number of entries and show that the large $N_c$ approximation for the KLWMIJ evolution coincides with the exact evolution \emph{i.e.} all the subleading  $N_c$ terms cancel in the action of $H_{KLWMIJ}$ on a single trace operator. We write the generic single trace operator as  
\beq
O_{2n}= \frac{1}{N_c}\tr \bigg(R_1R^{\dagger}_2...R_{2n-1}R^{\dagger}_{2n}\bigg)= \frac{1}{N_c}\tr\left[\prod_{i=1}^{n}R_{2i-1}R^{\dagger}_{2i}\right] .
\eeq
In order to find the action of $H_{KLWMIJ}$, we need the calculate the action of right and left rotation generators on $O_{2n}$. For $J_L$ we have
\beq
\label{Laction1}
J_L^a(y)O_{2n}=\frac{1}{N_c}\sum_{j=1}^n\tr\left[\left(\prod_{i=1}^{j-1}R_{2i-1}R^{\dagger}_{2i}\right)t^a\left(\prod_{i=j}^{n}R_{2i-1}R^{\dagger}_{2i}\right)\right]\delta(y_{2j-1}-y_{2j-2}) .
\eeq
Here, we introduced a short hand notation for the $\delta$-functions as
\beq
\delta(y_{2j-1}-y_{2j-2})\equiv\delta[y-(2j-1)]-\delta[y-(2j-2)] .
\eeq
Moreover, when writing  Eq.\eqref{Laction1} we have adopted the following conventions :
\beq
y_0&\equiv& y_{2n},\\
\prod_{i=k}^{l}R_{2i-1}R^{\dagger}_{2i}&\equiv&1\hspace{1cm} \text {for $ l < k$} .
\eeq
Similarly, one can write the action of $J_R$ on  $O_{2n}$ :
\beq
\label{Raction1}
\hspace{-0.5cm}J_R^a(y)O_{2n}=\frac{1}{N_c}\sum_{j=1}^n\tr\!\left[\left(\prod_{i=1}^{j-1}R_{2i-1}R^{\dagger}_{2i}\right)R_{2j-1}t^aR^{\dagger}_{2j}\left(\prod_{i=j+1}^{n}R_{2i-1}R^{\dagger}_{2i}\right)\right] \!\!\delta(y_{2j-1}-y_{2j}) .
\eeq
Now using Eqs.\eqref{Laction1} and \eqref{Raction1} we obtain
\beq
\label{LLaction}
&&\hspace{-1cm}J_L^a(x)J_L^a(y)O_{2n}=\frac{1}{N_c}\sum_{j=1}^n\tr\left[\left(\prod_{i=1}^{j-1}R_{2i-1}R^{\dagger}_{2i}\right)t^at^a\left(\prod_{i=j}^{n}R_{2i-1}R^{\dagger}_{2i}\right)\right]\nonumber\\
&&\hspace{3cm}\times\delta(y_{2j-1}-y_{2j-2})\delta(x_{2j-1}-x_{2j-2})\nonumber\\
&&\hspace{-0.5cm}+\frac{1}{N_c}\sum_{j=2}^n\sum_{k=1}^{j-1}\tr\left[\left(\prod_{i=1}^{k-1}R_{2i-1}R^{\dagger}_{2i}\right)t^a\left(\prod_{i=k}^{j-1}R_{2i-1}R^{\dagger}_{2i}\right)t^a\left(\prod_{i=j}^{n}R_{2i-1}R^{\dagger}_{2i}\right)\right]\nonumber\\
&&\times\left[ \delta(x_{2j-1}-x_{2j-2}) \delta(y_{2k-1}-y_{2k-2})+ \delta(y_{2j-1}-y_{2j-2}) \delta(x_{2k-1}-x_{2k-2})\right] ,
\eeq
\beq
\label{RRaction}
&&\hspace{-0.3cm}J_R^a(x)J_R^a(y)O_{2n}=\frac{1}{N_c}\sum_{j=1}^n\tr\left[\left(\prod_{i=1}^{j-1}R_{2i-1}R^{\dagger}_{2i}\right)R_{2j-1}t^at^aR^{\dagger}_{2j}\left(\prod_{i=j+1}^{n}R_{2i-1}R^{\dagger}_{2i}\right)\right]\nonumber\\
&&\hspace{3cm}\times\delta(y_{2j-1}-y_{2j})\delta(x_{2j-1}-x_{2j})\nonumber\\
&&\hspace{-0.3cm}+\frac{1}{N_c}\sum_{j=2}^n\sum_{k=1}^{j-1}\tr\!\!\left[\!\!\left(\prod_{i=1}^{k-1}\!\!R_{2i-1}R^{\dagger}_{2i}\right)\!\!R_{2k-1}t^aR^{\dagger}_{2k}\!\!\left(\prod_{i=k+1}^{j-1}\!\!\!R_{2i-1}R^{\dagger}_{2i}\right)R_{2j-1}t^aR^{\dagger}_{2j}\left(\prod_{i=j+1}^{n}\!\!\!R_{2i-1}R^{\dagger}_{2i}\right)\!\!\right]\nonumber\\
&&\hspace{2cm}\times\left[\delta(y_{2j-1}-y_{2j})\delta(x_{2k-1}-x_{2k})+\delta(y_{2k-1}-y_{2k})\delta(x_{2j-1}-x_{2j})\right],
\eeq
\beq
\label{LRaction}
&&\hspace{-0.5cm}J_L^a(x)J_R^b(y)O_{2n}=\nonumber\\
&&=\Bigg\{\frac{1}{N_c}\sum_{j=1}^n\sum_{k=1}^j\tr\left[\left(\prod_{i=1}^{k-1}R_{2i-1}R^{\dagger}_{2i}\right)t^a\left(\prod_{i=k}^{j-1}R_{2i-1}R^{\dagger}_{2i}\right)R_{2j-1}t^bR^{\dagger}_{2j}\left(\prod_{i=j+1}^{n}R_{2i-1}R^{\dagger}_{2i}\right)\right]\nonumber\\
&&+\frac{1}{N_c}\sum_{j=1}^n\sum_{k=j+1}^n\tr\left[\left(\prod_{i=1}^{j-1}R_{2i-1}R^{\dagger}_{2i}\right)R_{2j-1}t^bR^{\dagger}_{2j}\left(\prod_{i=j+1}^{k-1}R_{2i-1}R^{\dagger}_{2i}\right)t^a\left(\prod_{i=k}^{n}R_{2i-1}R^{\dagger}_{2i}\right)\right]\Bigg\}\nonumber\\
&&\hspace{5cm}\times\Big(\delta(y_{2j-1}-y_{2j})\delta(x_{2k-1}-x_{2k-2})\Big).
\eeq
These expressions can be simplified by using the color algebra relations 
\beq
\label{CA1}
&&\tr [At^at^aB]=\frac{N_c^2-1}{2N_c}\tr[AB] ,\\
\label{CA2}
&&\tr [At^aBt^aC]=\frac{1}{2}\left(\tr[AC]\tr[B]-\frac{1}{N_c}\tr[ABC]\right) .
\eeq 
Then Eqs. \eqref{CA1} and \eqref{CA2} can be rewritten as 

\beq
&&\hspace{-1cm}J_L^a(x)J_L^a(y)O_{2n}=\frac{1}{2}\tr\left[\prod_{i=1}^{n}R_{2i-1}R^{\dagger}_{2i}\right]\sum_{j=1}^n\delta(y_{2j-1}-y_{2j-2})\delta(x_{2j-1}-x_{2j-2})\nonumber\\
&&+\frac{1}{2N_c}\sum_{j=2}^n\sum_{k=1}^{j-1}\tr\left[\left(\prod_{i=1}^{k-1}R_{2i-1}R^{\dagger}_{2i}\right)\left(\prod_{i=j}^{n}R_{2i-1}R^{\dagger}_{2i}\right)\right]\tr\left[\prod_{i=k}^{j-1}R_{2i-1}R^{\dagger}_{2i}\right]\nonumber\\
&&\hspace{1cm}\times\Big(\delta(y_{2j-1}-y_{2j-2})\delta(x_{2k-1}-x_{2k-2})+\delta(y_{2k-1}-y_{2k-2})\delta(x_{2j-1}-x_{2j-2})\Big)\nonumber\\
&&-\frac{1}{2N^2_c}\tr\left[\prod_{i=1}^{n}R_{2i-1}R^{\dagger}_{2i}\right]\Bigg\{\sum_{j=1}^n\delta(y_{2j-1}-y_{2j-2})\delta(x_{2j-1}-x_{2j-2})\nonumber\\
&&\hspace{-0.8cm}+\sum_{j=2}^n\sum_{k=1}^{j-1}\Big(\delta(y_{2j-1}-y_{2j-2})\delta(x_{2k-1}-x_{2k-2})+\delta(y_{2k-1}-y_{2k-2})\delta(x_{2j-1}-x_{2j-2})\Big)
\Bigg\} .
\eeq
Note that the $N_c$ suppressed terms can be further simplified by using the identity
\beq
\sum_{j=1}^na_jb_j+\sum_{j=2}^n\sum_{k=1}^{j-1}\big(b_ja_k+b_ka_j\big)=\sum_{j=1}^na_j\sum_{k=1}^nb_k .
\eeq
The final expression for $J_L^a(x)J_L^a(y)O_{2n}$ is
\beq
\label{LLcomp}
&&\hspace{-0.7cm}J_L^a(x)J_L^a(y)O_{2n}=\frac{1}{2}\tr\left[\prod_{i=1}^{n}R_{2i-1}R^{\dagger}_{2i}\right]\Bigg\{\sum_{j=1}^n\delta(y_{2j-1}-y_{2j-2})\delta(x_{2j-1}-x_{2j-2})\nonumber\\
&&\hspace{6cm}-\frac{1}{N^2_c}\sum_{j=1}^n\delta(y_{2j-1}-y_{2j-2})\sum_{k=1}^n\delta(x_{2k-1}-x_{2k-2})\Bigg\}\nonumber\\
&&+\frac{1}{2N_c}\sum_{j=2}^n\sum_{k=1}^{j-1}\tr\left[\left(\prod_{i=1}^{k-1}R_{2i-1}R^{\dagger}_{2i}\right)\left(\prod_{i=j}^{n}R_{2i-1}R^{\dagger}_{2i}\right)\right]\tr\left[\prod_{i=k}^{j-1}R_{2i-1}R^{\dagger}_{2i}\right]\nonumber\\
&&\hspace{1cm}\times\Big(\delta(y_{2j-1}-y_{2j-2})\delta(x_{2k-1}-x_{2k-2})+\delta(y_{2k-1}-y_{2k-2})\delta(x_{2j-1}-x_{2j-2})\Big).
\eeq
Similarly,
\beq
\label{RRcomp}
&&\hspace{-0.4cm}J_R^a(x)J_R^a(y)O_{2n}=\frac{1}{2}\tr\left[\prod_{i=1}^{n}R_{2i-1}R^{\dagger}_{2i}\right]\Bigg\{\sum_{j=1}^n\delta(y_{2j-1}-y_{2j})\delta(x_{2j-1}-x_{2j})\nonumber\\
&&\hspace{6.7cm}-\frac{1}{N^2_c}\sum_{j=1}^n\delta(y_{2j-1}-y_{2j})\sum_{k=1}^n\delta(x_{2k-1}-x_{2k})\Bigg\}\nonumber\\
&&\hspace{-0.4cm}+\frac{1}{2N_c}\sum_{j=2}^n\sum_{k=1}^{j-1}\tr\!\!\left[\!\!\left(\prod_{i=1}^{k-1}\!R_{2i-1}R^{\dagger}_{2i}\!\!\right)\!\!R_{2k-1}R^{\dagger}_{2j}\!\!\left(\prod_{i=j+1}^{n}\!\!\!R_{2i-1}R^{\dagger}_{2i}\!\!\right)\!\!\right]\!\tr\!\!\left[\!R_{2j-1}R^{\dagger}_{2k}\!\!\left(\prod_{i=k+1}^{j-1}\!\!\!R_{2i-1}R^{\dagger}_{2i}\!\right)\!\right]\nonumber\\
&&\hspace{1cm}\times\Big(\delta(y_{2j-1}-y_{2j})\delta(x_{2k-1}-x_{2k})+\delta(y_{2k-1}-y_{2k})\delta(x_{2j-1}-x_{2j})\Big).
\eeq
The real term in the KLWMIJ Hamiltonian contains a factor of $(-2R^{ab}_z)$ where the unitary matrix $R^{ab}_z$ is in the adjoint representation. Thus, the color algebra is slightly different for this term. Again using the completeness relation and writing the adjoint matrix in terms of the product of two fundamental ones we get
\beq
\label{cAwR}
-2R^{ab}_z\tr[At^aBt^bC]=-\left\{\tr[R_zCA]\tr[R^{\dagger}_zB]-\frac{1}{N_c}\tr[ABC]\right\} .
\eeq
Using Eq.\eqref{cAwR},the real term can be written as 
\beq
\label{LRcomp}
&&\hspace{-0.3cm}
-2J_L^a(x)R^{ab}_zJ_R^b(y)O_{2n}=\frac{1}{N^2_c}\tr\left[\prod_{i=1}^{n}R_{2i-1}R^{\dagger}_{2i}\right]\sum_{j=1}^n\delta(y_{2j-1}-y_{2j})\sum_{k=1}^n\delta(x_{2k-1}-x_{2k})\nonumber\\
&&\hspace{-0.3cm}
-\frac{1}{N_c}\Bigg\{
\sum_{j=1}^{n}\sum_{k=1}^{j} 
\tr \!\! \left[ \left(\prod_{i=1}^{k-1}R_{2i-1}R^{\dagger}_{2i}\right) \!\! R_zR^{\dagger}_{2j} \!\! \left(\prod_{i=j+1}^{n}R_{2i-1}R^{\dagger}_{2i}\right) \right]
\tr \!\! \left[ \left(\prod_{i=k}^{j-1}R_{2i-1}R^{\dagger}_{2i}\right)R_{2j-1}R^{\dagger}_z \right] \nonumber\\
&&\hspace{-0.3cm}
+ \sum_{j=1}^{n}\sum_{k=j+1}^{n} 
\tr \!\! \left[ R_z R^{\dagger}_{2j}  \left(\prod_{i=j+1}^{k-1}R_{2i-1}R^{\dagger}_{2i}\right) \right] 
\tr \!\! \left[ \left(\prod_{i=1}^{j-1}R_{2i-1}R^{\dagger}_{2i}\right) \!\! R_{2j-1}R^{\dagger}_z  \!\! \left(\prod_{i=k}^{n}R_{2i-1}R^{\dagger}_{2i}\right) \right]
\Bigg\}\nonumber\\
&&\hspace{8cm}\times \delta(y_{2j-1}-y_{2j})\delta(x_{2k-1}-x_{2k-2}).
\eeq
Now, combining all the pieces Eqs.\eqref{LLcomp}, \eqref{RRcomp} and \eqref{LRcomp}, one obtains
\beq
\label{compEvolO2n}
&&H_{KLWMIJ}O_{2n}=\frac{\bar{\alpha_s}}{2\pi}\int_{xyz}-\frac{1}{2}M_{xyz}
\Bigg\{\frac{1}{2N_c}\tr\left[\prod_{i=1}^{n}R_{2i-1}R^{\dagger}_{2i}\right]\nonumber\\
&&\hspace{1cm}\times
\sum_{j=1}^n\Big(\delta(y_{2j-1}-y_{2j})\delta(x_{2j-1}-x_{2j})+\delta(y_{2j-1}-y_{2j-2})\delta(x_{2j-1}-x_{2j-2})\Big)\nonumber\\
&&+\frac{1}{N^2_c}\sum_{j=2}^n\sum_{k=1}^{j-1}\tr\left[\left(\prod_{i=1}^{k-1}R_{2i-1}R^{\dagger}_{2i}\right)\left(\prod_{i=j}^{n}R_{2i-1}R^{\dagger}_{2i}\right)\right]\tr\left[\prod_{i=k}^{j-1}R_{2i-1}R^{\dagger}_{2i}\right]\nonumber\\
&&\hspace{8cm}\times\delta(y_{2j-1}-y_{2j-2})\delta(x_{2k-1}-x_{2k-2})\nonumber\\
&&\hspace{-0.4cm}+\frac{1}{N^2_c}\sum_{j=2}^n\sum_{k=1}^{j-1}\tr\!\!\left[\!\!\left(\prod_{i=1}^{k-1}\!R_{2i-1}R^{\dagger}_{2i}\!\!\right)\!\!R_{2k-1}R^{\dagger}_{2j}\!\!\left(\prod_{i=j+1}^{n}\!\!\!R_{2i-1}R^{\dagger}_{2i}\!\!\right)\!\!\right]\!\tr\!\!\left[\!R_{2j-1}R^{\dagger}_{2k}\!\!\left(\prod_{i=k+1}^{j-1}\!\!\!R_{2i-1}R^{\dagger}_{2i}\!\right)\!\right]\nonumber\\
&&\hspace{8cm}\times\delta(y_{2j-1}-y_{2j})\delta(x_{2k-1}-x_{2k})\nonumber\\
&&\hspace{-0.3cm}
-\frac{1}{N^2_c}
\sum_{j=1}^{n}\sum_{k=1}^{j} 
\tr \!\! \left[ \left(\prod_{i=1}^{k-1}R_{2i-1}R^{\dagger}_{2i}\right) \!\! R_zR^{\dagger}_{2j} \!\! \left(\prod_{i=j+1}^{n}R_{2i-1}R^{\dagger}_{2i}\right) \right]
\tr \!\! \left[ \left(\prod_{i=k}^{j-1}R_{2i-1}R^{\dagger}_{2i}\right)R_{2j-1}R^{\dagger}_z \right] \nonumber\\
&&\hspace{8cm}\times \delta(y_{2j-1}-y_{2j})\delta(x_{2k-1}-x_{2k-2})\nonumber\\
&&\hspace{-0.3cm}
-\frac{1}{N^2_c}
\sum_{j=1}^{n}\sum_{k=j+1}^{n} 
\tr \!\! \left[ R_z R^{\dagger}_{2j}  \left(\prod_{i=j+1}^{k-1}R_{2i-1}R^{\dagger}_{2i}\right) \right] 
\tr \!\! \left[ \left(\prod_{i=1}^{j-1}R_{2i-1}R^{\dagger}_{2i}\right) \!\! R_{2j-1}R^{\dagger}_z  \!\! \left(\prod_{i=k}^{n}R_{2i-1}R^{\dagger}_{2i}\right) \right]
\nonumber\\
&&\hspace{7.8cm}\times \delta(y_{2j-1}-y_{2j})\delta(x_{2k-1}-x_{2k-2})\Bigg\}.
\eeq
Eq.\eqref{compEvolO2n} gives the complete KLWMIJ evolution of a generic single trace operator and it is equivalent to Eq.\eqref{ho2n}. All the terms subleading in $N_c$ cancel in this equation. 

\subsection{KLWMIJ evolution of a generic double trace operator}
We now derive the evolution equation of a double trace operator.
Consider two generic operators $O_{2n}$ and $O_{2\bar{m}}$ : 
\beq
&&O_{2n}= \frac{1}{N_c}\tr(R_1R^{\dagger}_2...R_{2n-1}R^{\dagger}_{2n})= \frac{1}{N_c}\tr\left[\prod_{i=1}^{n}R_{2i-1}R^{\dagger}_{2i}\right],\\
&&O_{2\bar{m}}= \frac{1}{N_c}\tr(R_{\bar{1}}R^{\dagger}_{\bar{2}}...R_{2\bar{m}-\bar{1}}R^{\dagger}_{2\bar{m}})= \frac{1}{N_c}\tr\left[\prod_{i=\bar{1}}^{\bar{m}}R_{2i-\bar{1}}R^{\dagger}_{2i}\right].
\eeq 

In the evolution of the product of these two objects, the two $SU(N_c)$ rotation generators appearing in $H_{KLWMIJ}$ can either act on the same object keeping the other one untouched, or  one rotation generator can act on one object and the second one on the other object. Therefore, generically 
\beq
\label{genEvl}
H_{KLWMIJ}O_{2n}O_{2\bar{m}} = \left[H_{KLWMIJ}O_{2n}\right]O_{2\bar{m}}  +  O_{2n} \left[H_{KLWMIJ}O_{2\bar{m}}\right]  + \chi_{\rm{mix}},
\eeq
with
\beq
\label{chimix}
&&\hspace{-1.4cm}
\chi_{\rm{mix}}=\frac{\alpha_s}{2\pi^2}\int_{xyz} \!\!\!\!-\frac{1}{2}M_{xyz}
\Bigg\{
\left[ J_L^a(x) O_{2n}\right] \left[ J_L^a(y) O_{2\bar{m}}\right]  
+ \left[ J_R^a(x) O_{2n}\right]  \left[ J_R^a(y) O_{2\bar{m}}\right] \nonumber\\
&&\hspace{2.2cm}
- \left[ J_L^a(x) O_{2n}\right]  R^{ab}_z \left[ J_R^b(y) O_{2\bar{m}}\right] 
- \left[ J_L^a(x) O_{2\bar{m}}\right]  R^{ab}_z \left[ J_R^b(y) O_{2n}\right]
\Bigg\}.
\eeq 
 By using  Eq.\eqref{Laction}, the first term in Eq.\eqref{chimix} can be written as
\beq
\label{LL}
&&\hspace{-0.4cm}
\left[ J_L^a(x) O_{2n}\right] \left[ J_L^a(y) O_{2\bar{m}}\right]  = \frac{1}{N^2_c} 
\sum_{j=1}^{n} 
\tr \left[ \left(\prod_{i=1}^{j-1}R_{2i-1}R^{\dagger}_{2i}\right) t^a \left(\prod_{i=j}^{n}R_{2i-1}R^{\dagger}_{2i}\right) \right] 
\delta(x_{2j-1}-x_{2j-2})\nonumber\\
&&\hspace{3cm}
\times
\sum_{k=\bar{1}}^{\bar{m}} 
\tr\left[ \left(\prod_{i=\bar{1}}^{k-\bar{1}}R_{2i-\bar{1}}R^{\dagger}_{2i}\right) t^a \left(\prod_{i=k}^{\bar{m}}R_{2i-\bar{1}}R^{\dagger}_{2i}\right) \right]
\delta(y_{2k-\bar{1}}-y_{2k-\bar{2}}),
\eeq
Further, using 
\beq
\label{color2tr}
\tr[At^aB] \tr[Ct^aD] = \frac{1}{2} \left\{ \tr[BADC] - \frac{1}{N_c} \tr[AB]  \tr[CD] \right\},
\eeq
we have
\beq
\label{LLfinal}
&&
\left[ J_L^a(x) O_{2n}\right] \left[ J_L^a(y) O_{2\bar{m}}\right]  = \nonumber\\
&&
= \frac{1}{2N^2_c} 
\sum_{j=1}^{n} \sum_{k=\bar{1}}^{\bar{m}} \Bigg\{
\tr \left[
\left(\prod_{i=j}^{n}R_{2i-1}R^{\dagger}_{2i}\right) \left(\prod_{i=1}^{j-1}R_{2i-1}R^{\dagger}_{2i}\right) 
\left(\prod_{i=k}^{\bar{m}}R_{2i-\bar{1}}R^{\dagger}_{2i}\right) \left(\prod_{i=\bar{1}}^{k-\bar{1}}R_{2i-\bar{1}}R^{\dagger}_{2i}\right)
\right]
\nonumber\\
&&\hspace{1cm}
-\frac{1}{N_c} \tr\left[\prod_{i=1}^{n}R_{2i-1}R^{\dagger}_{2i}\right] \tr\left[\prod_{i=\bar{1}}^{\bar{m}}R_{2i-\bar{1}}R^{\dagger}_{2i}\right] \Bigg\}
\delta(y_{2k-\bar{1}}-y_{2k-\bar{2}})
\delta(x_{2j-1}-x_{2j-2}).
\eeq
By using Eq.\eqref{Raction} the action of the virtual term can be written as
\beq
\label{RR}
&&\hspace{-1.3cm}
\left[ J_R^a(x) O_{2n}\right] \left[ J_R^a(y) O_{2\bar{m}}\right] = \nonumber\\
&& 
=\frac{1}{N^2_c} 
\sum_{j=1}^{n} 
\tr \left[ \left(\prod_{i=1}^{j-1}R_{2i-1}R^{\dagger}_{2i}\right)R_{2j-1} t^a R^{\dagger}_{2j}\left(\prod_{i=j+1}^{n}R_{2i-1}R^{\dagger}_{2i}\right) \right] 
\delta(x_{2j-1}-x_{2j})\nonumber\\
&&\hspace{0.5cm}
\times
\sum_{k=\bar{1}}^{\bar{m}} 
\tr \left[ \left(\prod_{i=\bar{1}}^{k-\bar{1}}R_{2i-\bar{1}}R^{\dagger}_{2i}\right)R_{2k-\bar{1}} t^a R^{\dagger}_{2k}\left(\prod_{i=k+\bar{1}}^{\bar{m}}R_{2i-\bar{1}}R^{\dagger}_{2i}\right) \right] 
\delta(y_{2k-\bar{1}}-y_{2k}).
\eeq
Performing the color algebra we get
\beq
\label{RRfinal}
&&\hspace{-0.3cm}
\left[ J_R^a(x) O_{2n}\right] \left[ J_R^a(y) O_{2\bar{m}}\right] = 
\frac{1}{2N^2_c} 
\sum_{j=1}^{n} \sum_{k=\bar{1}}^{\bar{m}}\Bigg\{\nonumber\\
&&\hspace{-0.3cm}\times \!
\tr \!\!  \left[ \!R^{\dagger}_{2j} \!\! \left(\prod_{i=j+1}^{n} \!\! R_{2i-1}R^{\dagger}_{2i}\!\right)  \!\! \left(\prod_{i=1}^{j-1}R_{2i-1}R^{\dagger}_{2i}\!\right) \!\!R_{2j-1}  R^{\dagger}_{2k}\!\!\left(\prod_{i=k+\bar{1}}^{\bar{m}}\!\!R_{2i-\bar{1}}R^{\dagger}_{2i}\!\right) \!\!\!\left(\prod_{i=\bar{1}}^{k-\bar{1}}R_{2i-\bar{1}}R^{\dagger}_{2i}\!\right) \!\!R_{2k-\bar{1}}\!\right]  \nonumber\\
&&
-\frac{1}{N_c} \tr\left[\prod_{i=1}^{n}R_{2i-1}R^{\dagger}_{2i}\right] \tr\left[\prod_{i=\bar{1}}^{\bar{m}}R_{2i-\bar{1}}R^{\dagger}_{2i}\right] \Bigg\}
\delta(y_{2k-\bar{1}}-y_{2k})
\delta(x_{2j-1}-x_{2j}) .
\eeq
The real term in $H_{KLWMIJ}$ contributes to the ``mixed'' term:
\beq
\label{MixTerm1}
&&\hspace{-0.3cm}
-R_z^{ab}\Big[ J_L^a(x) O_{2n}\Big] \left[ J_R^b(y) O_{2\bar{m}}\right] \!=\! -\frac{R_z^{ab}}{N^2_c} 
\sum_{j=1}^{n}\tr \left[ \left(\prod_{i=1}^{j-1}R_{2i-1}R^{\dagger}_{2i}\right) t^a \left(\prod_{i=j}^{n}R_{2i-1}R^{\dagger}_{2i}\right) \right] \\
&&\times\delta(x_{2j-1}-x_{2j-2})
%
 \sum_{k=\bar{1}}^{\bar{m}}\tr \!\left[ \!\left(\prod_{i=\bar{1}}^{k-\bar{1}}R_{2i-\bar{1}}R^{\dagger}_{2i}\right)\!R_{2k-\bar{1}} t^bR^{\dagger}_{2k}\!\left(\prod_{i=k+\bar{1}}^{m}R_{2i-\bar{1}}R^{\dagger}_{2i}\!\right)\! \right] \!\delta(y_{2k-\bar{1}}\!-\!y_{2k}).\nonumber
\eeq
The relevant piece of color algebra here is
\beq
\label{colorMix1}
-R_z^{ab}\tr[At^aB]\tr[Ct^bD]=-\frac{1}{2}\left(\tr[R^{\dagger}_zBAR_zDC]-\frac{1}{N_c}\tr[AB]\tr[CD]\right).
\eeq
Using Eq.\eqref{colorMix1}, the real ``mixed'' term, Eq.\eqref{MixTerm1}, can be written as 
\beq
&&-R_z^{ab}\Big[ J_L^a(x) O_{2n}\Big] \left[ J_R^b(y) O_{2m}\right] \!=\!
-\frac{1}{2N^2_c}\sum_{j=1}^{n}\sum_{k=\bar{1}}^{\bar{m}}\bigg\{\!
\tr\!\Bigg[\!R^{\dagger}_z \left(\prod_{i=j}^{n}R_{2i-1}R^{\dagger}_{2i}\right)  \left(\prod_{i=1}^{n}R_{2i-1}R^{\dagger}_{2i}\right)\nonumber\\
&&
\hspace{5cm}
\times
R_zR^{\dagger}_{2k}  \left(\prod_{i=k+\bar{1}}^{\bar{m}}R_{2i-\bar{1}}R^{\dagger}_{2i}\right) \left(\prod_{i=\bar{1}}^{k-\bar{1}}R_{2i-\bar{1}}R^{\dagger}_{2i}\right)R_{2k-\bar{1}} \Bigg]\nonumber\\
&&\hspace{0.8cm}
-\frac{1}{N_c}\tr\left[\prod_{i=1}^{n}R_{2i-1}R^{\dagger}_{2i} \right] \tr\left[\prod_{i=\bar{1}}^{\bar{m}}R_{2i-\bar{1}}R^{\dagger}_{2i} \right]\Bigg\}\delta(x_{2j-1}-x_{2j-2})\delta(y_{2k-\bar{1}}-y_{2k}).
\eeq
Similarly,
\beq
\label{MixTerm2}
&&\hspace{-0.3cm}
-R_z^{ab}\Big[ J_L^a(x) O_{2\bar{m}}\Big] \left[ J_R^b(y) O_{2n}\right] \!=\! -\frac{R_z^{ab}}{N^2_c} 
\sum_{k=\bar{1}}^{\bar{m}}\tr \left[ \left(\prod_{i=\bar{1}}^{k-\bar{1}}R_{2i-\bar{1}}R^{\dagger}_{2i}\right) t^a \left(\prod_{i=k}^{\bar{m}}R_{2i-\bar{1}}R^{\dagger}_{2i}\right) \right] \nonumber\\
&&\hspace{1.5cm}\times
 \sum_{j=1}^{n}\tr \left[ \left(\prod_{i=1}^{j-1}R_{2i-1}R^{\dagger}_{2i}\right)R_{2j-1} t^b R^{\dagger}_{2j}\left(\prod_{i=j+1}^{n}R_{2i-1}R^{\dagger}_{2i}\right) \right] \nonumber\\
 &&\hspace{1.5cm}\times\delta(x_{2k-\bar{1}}-x_{2k-\bar{2}})
\delta(y_{2j-1}-y_{2j})
\eeq
can be written as
\beq
\label{MixTerm2Final}
&&-R_z^{ab}\Big[ J_L^a(x) O_{2\bar{m}}\Big] \left[ J_R^b(y) O_{2n}\right] =
-\frac{1}{2N^2_c}\sum_{j=1}^{n}\sum_{k=\bar{1}}^{\bar{m}}\bigg\{\tr\Bigg[R^{\dagger}_z \left(\prod_{i=k}^{\bar{m}}R_{2i-\bar{1}}R^{\dagger}_{2i}\right)  \left(\prod_{i=\bar{1}}^{k-\bar{1}}R_{2i-\bar{1}}R^{\dagger}_{2i}\right)\nonumber\\
&&\hspace{6cm}\times
R_zR^{\dagger}_{2j}  \left(\prod_{i=j+1}^{n}R_{2i-1}R^{\dagger}_{2i}\right) \left(\prod_{i=1}^{j-1}R_{2i-1}R^{\dagger}_{2i}\right)R_{2j-1} \Bigg]\nonumber\\
&&\hspace{1cm}
-\frac{1}{N_c}\tr\left[\prod_{i=1}^{n}R_{2i-1}R^{\dagger}_{2i} \right] \tr\left[\prod_{i=\bar{1}}^{\bar{m}}R_{2i-\bar{1}}R^{\dagger}_{2i} \right]\Bigg\}\delta(x_{2k-\bar{1}}-x_{2k-\bar{2}})\delta(y_{2j-1}-y_{2j}).
\eeq
Combining all the terms we see that all the order $1/N^3_c$ terms cancel. Hence,
\beq
\label{DoubleTraceEvolution}
&&\hspace{-0.25cm}
\chi_{\rm{mix}} = \frac{\alpha_s}{2\pi^2}\int_{xyz} \!\!\!\!-\frac{1}{2}M_{xyz}
\frac{1}{2N^2_c}\sum_{j=1}^{n}\sum_{k=\bar{1}}^{\bar{m}}
\Bigg\{\nonumber\\
&&\hspace{-0.25cm}\times
\tr\!\!\left[  \left(\prod_{i=j}^{n}R_{2i-1}R^{\dagger}_{2i}\right)  \left(\prod_{i=1}^{j-1}R_{2i-1}R^{\dagger}_{2i}\right)   \left(\prod_{i=k}^{\bar{m}}R_{2i-\bar{1}}R^{\dagger}_{2i}\right) 
 \left(\prod_{i=\bar{1}}^{k-\bar{1}}R_{2i-\bar{1}}R^{\dagger}_{2i}\right) \right] \nonumber\\
 &&\hspace{3cm}
 \times
 \delta(x_{2j-1}-x_{2j-2})\delta(y_{2k-\bar{1}}-y_{2k-\bar{2}})\nonumber\\
&&\hspace{-0.25cm}+
\tr\!\!\left[\! R^{\dagger}_{2j} \!\!\left(\prod_{i=j+1}^{n}\!\!\!\!R_{2i-1}R^{\dagger}_{2i}\!\!\right) \!\! \left(\prod_{i=1}^{j-1}\!\!R_{2i-1}R^{\dagger}_{2i}\!\!\right) \!\!R_{2j-1} R^{\dagger}_{2k}\!\! \left(\prod_{i=k+\bar{1}}^{\bar{m}}\!\!\!\!R_{2i-\bar{1}}R^{\dagger}_{2i}\!\!\right)\!\! 
 \left(\prod_{i=\bar{1}}^{k-\bar{1}}R_{2i-\bar{1}}R^{\dagger}_{2i}\!\!\right) \!\!R_{2k-\bar{1}}\right]\nonumber\\
 &&\hspace{3cm}
  \times
 \delta(x_{2j-1}-x_{2j})\delta(y_{2k-\bar{1}}-y_{2k})\nonumber\\
&&\hspace{-0.25cm}-
\tr\!\!\left[\! R^{\dagger}_{z}\!\! \left(\prod_{i=j}^{n}R_{2i-1}R^{\dagger}_{2i}\!\!\right) \!\! \left(\prod_{i=1}^{j-1}R_{2i-1}R^{\dagger}_{2i}\!\!\right)\!\! R_{z} R^{\dagger}_{2k} \!\!\left(\prod_{i=k+\bar{1}}^{\bar{m}}\!\!R_{2i-\bar{1}}R^{\dagger}_{2i}\!\!\right)\!\! 
 \left(\prod_{i=\bar{1}}^{k-\bar{1}}R_{2i-\bar{1}}R^{\dagger}_{2i}\!\!\right) \!\!R_{2k-\bar{1}}\right]\nonumber\\
 && \hspace{3cm}
 \times
 \delta(x_{2j-1}-x_{2j-2})\delta(y_{2k-\bar{1}}-y_{2k})\nonumber\\
&&\hspace{-0.25cm}-
\tr\!\!\left[\! R^{\dagger}_{z}\!\! \left(\prod_{i=k}^{\bar{m}}\!\!R_{2i-\bar{1}}R^{\dagger}_{2i}\!\!\right) \!\! \left(\prod_{i=\bar{1}}^{k-\bar{1}}\!\!R_{2i-\bar{1}}R^{\dagger}_{2i}\!\!\right) \!\!R_{z} R^{\dagger}_{2j} \!\!\left(\prod_{i=j+1}^{n}\!\!R_{2i-1}R^{\dagger}_{2i}\!\!\right) \!\!
 \left(\prod_{i=1}^{j-1}\!\!R_{2i-1}R^{\dagger}_{2i}\!\!\right) \!\!R_{2j-1}\right]\nonumber\\
 &&\hspace{3cm}
  \times
 \delta(x_{2k-\bar{1}}-x_{2k-\bar{2}})\delta(y_{2j-1}-y_{2j})\Bigg\}.
\eeq
The delta functions can be now realized performing integrations over $x$ and $y$.
Combining Eqs.\eqref{compEvolO2n} and  \eqref{DoubleTraceEvolution}, we arrive at Eq.\eqref{genralized_decomposedH}. 

\section{Appendix 2: Evolution of mutipoles.}

The evolution equation of a quadrupole is \cite{KLW, JK}
\beq
\label{Qev}
&&\hspace{-1.5cm}\frac{d}{dY}Q_{1234}=\frac{\bar{\alpha_s}}{2\pi}\int_z -\frac{1}{2}\big(M_{12z}\!+\!M_{23z}\!+\!M_{34z}\!+\!M_{14z}\big)Q_{1234}\!-\!L_{1234z}d_{14}d_{32}\!-\!L_{1432z}d_{12}d_{34}\nonumber\\
&&\hspace{1.2cm}\!+\!L_{1214z}d_{1z}Q_{z234}\!+\!L_{1232z}d_{z2}Q_{1z34}\!+\!L_{3234z}d_{3z}Q_{12z4}\!+\!L_{1434z}d_{z4}Q_{123z}.
\eeq
Using the definitions given in Eqs.\eqref{4BR}, \eqref{4CR}, \eqref{4DM} and \eqref{4DP} one can calculate the evolution equation of 4-point Reggeons \cite{last}
\beq
&&\hspace{-2cm}\frac{d}{dY}Q^{++}_{1234}=\frac{\bar{\alpha_s}}{2\pi}\int_z-\frac{1}{2}\big(M_{12z}\!+\!M_{23z}\!+\!M_{34z}\!+\!M_{14z}\big)Q^{++}_{1234}\nonumber\\
&&\hspace{1cm}+L_{1234z}\big[(P_{1z}\!-\!P_{4z})(P_{3z}\!-\!P_{2z})\!+\!P_{14}P_{32}\!+\!O_{14}O_{32}\big]\nonumber\\
&&\hspace{1cm}+L_{1432z}\big[(P_{1z}\!-\!P_{2z})(P_{3z}\!-\!P_{z4})\!+\!P_{12}P_{34}\!+\!O_{12}O_{34}\big]\nonumber\\
&&\hspace{1cm}+L_{1214z}\big[Q^{++}_{z234}(1\!-\!P_{1z})\!-\!Q^{--}_{z234}O_{1z}\!-\!P_{1z}(P_{23}\!-\!P_{24}\!+\!P_{34}) \big]\nonumber\\
&&\hspace{1cm}+L_{1232z}\big[Q^{++}_{1z34}(1\!-\!P_{z2})\!-\!Q^{--}_{1z34}O_{z2}\!-\!P_{z2}(P_{14}\!-\!P_{13}\!+\!P_{34}) \big]\nonumber\\
&&\hspace{1cm}+L_{3234z}\big[Q^{++}_{12z4}(1\!-\!P_{3z})\!-\!Q^{--}_{12z4}O_{3z}\!-\!P_{3z}(P_{12}\!-\!P_{24}\!+\!P_{14}) \big]\nonumber\\
&&\hspace{1cm}+L_{1434z}\big[Q^{++}_{123z}(1\!-\!P_{z4})\!-\!Q^{--}_{123z}O_{z4}\!-\!P_{z4}(P_{12}\!-\!P_{13}\!+\!P_{23}) \big],
\eeq
\beq
\label{QCev}
&&\hspace{-1.2cm}\frac{d}{dY}Q^{+-}_{1234}=\frac{\bar{\alpha_s}}{2\pi}\int_z -\frac{1}{2}\big(M_{12z}\!+\!M_{23z}\!+\!M_{34z}\!+\!M_{14z}\big)Q^{+-}_{1234}\nonumber\\
&&\hspace{0.8cm}+L_{1214z}\big[Q^{+-}_{z234}(1-P_{1z})+Q^{-+}_{z234}O_{1z}\big]+L_{1232z}\big[Q^{+-}_{1z34}(1-P_{z2})+Q^{-+}_{1z34}O_{z2}\big]
\nonumber\\
&&\hspace{0.8cm}+L_{3234z}\big[Q^{+-}_{12z4}(1-P_{3z})+Q^{-+}_{12z4}O_{3z}\big]+L_{1434z}\big[Q^{+-}_{123z}(1-P_{z4})-Q^{-+}_{123z}O_{z4}\big].\nonumber\\
\eeq

The evolution equation of a 6-point function can be calculated by taking the action of the KLWMIJ Hamiltonian. The resulting expression is 
\beq
\label{Xev}
&&\hspace{-1.8cm}\frac{d}{dY}X_{123456}=\frac{\bar{\alpha_s}}{2\pi}\int_z -\frac{1}{2}\big(M_{12z}+M_{34z}+M_{56z}+M_{16z}+M_{23z}+M_{45z}\big)X_{123456}\nonumber\\
&&\hspace{2cm}+L_{1623z}d_{12}Q_{3456}+L_{2354z}d_{34}Q_{1256}+L_{1645z}d_{56}Q_{1234}\nonumber\\
&&\hspace{2cm}+L_{1243z}d_{32}Q_{1456}+L_{1265z}d_{16}Q_{5234}+L_{3465z}d_{54}Q_{1236}\nonumber\\
&&\hspace{2cm}-L_{1245z}Q_{1z56}Q_{z234}-L_{1643z}Q_{123z}Q_{z456}-L_{2356z}Q_{12z6}Q_{345z}\nonumber\\
&&\hspace{2cm}+L_{1216z}d_{1z}X_{z23456}+L_{1232z}d_{z2}X_{1z3456}+L_{2343z}d_{3z}X_{12z456}\nonumber\\
&&\hspace{2cm}+L_{3454z}d_{z4}X_{123z56}+L_{4565z}d_{5z}X_{1234z6}+L_{1656z}d_{z6}X_{12345z}.
\eeq

In the evolution of the 6-point function the terms with six points are either the hexapole or the breaking of the hexapole into two lower color singlet states which in this case it is a dipole and a quadrupole. The second structure which introduces two more points, breaks into two possible lower color singlet states; either a dipole and a hexapole or two quadrupoles. This is the same trend that was seen in the evolution of the 4-point function.

By using Eq.\eqref{XB}, one can write the evolution equation of the 6-point $++$-Reggeon as
\beq
&&\frac{d}{dY}X^{++}_{123456}=\frac{\bar{\alpha_s}}{2\pi}\int_z-\frac{1}{2}\big[ M_{12z}+M_{23z}+M_{34z}+M_{45z}+M_{56z}+M_{61z}\big]X^{++}_{123456}\nonumber\\
&&+L_{1643z}\big[(P_{12}\!-\!P_{13}\!+\!P_{23})Q^{++}_{z456}\!+\!(P_{45}\!-\!P_{46}\!+\!P_{56})Q^{++}_{123z}\!+\!\bar{Q}_{123z}\bar{Q}_{z456}\big]\nonumber\\
&&+L_{3265z}\big[(P_{34}\!-\!P_{35}\!+\!P_{45})Q^{++}_{12z6}\!+\!(P_{12}\!-\!P_{26}\!+\!P_{16})Q^{++}_{z345}\!+\!\bar{Q}_{12z6}\bar{Q}_{z345}\big]\nonumber\\
&&+L_{1245z}\big[(P_{16}\!-\!P_{15}\!+\!P_{56})Q^{++}_{z234}\!+\!(P_{23}\!-\!P_{24}\!+\!P_{34})Q^{++}_{1z56}\!+\!\bar{Q}_{z234}\bar{Q}_{1z56}\big]\nonumber\\
%
&&+L_{1216z}\big\{(1\!-\!P_{1z})X^{++}_{z23456}\!-\!O_{1z}X^{--}_{z23456}\!-\!P_{1z}\big[Q^{++}_{2345}\!-\!Q^{++}_{2346}\!+\!Q^{++}_{2356}\!-\!Q^{++}_{2456}\!+\!Q^{++}_{3456}\big]\big\}\nonumber\\
&&+L_{2123z}\big\{(1\!-\!P_{z2})X^{++}_{1z3456}\!-\!O_{z2}X^{--}_{1z3456}\!-\!P_{z2}\big[Q^{++}_{1346}\!-\!Q^{++}_{1345}\!+\!Q^{++}_{1456}\!-\!Q^{++}_{1356}\!+\!Q^{++}_{3456}\big]\big\}\nonumber\\
&&+L_{2343z}\big\{(1\!-\!P_{3z})X^{++}_{12z456}\!-\!O_{3z}X^{--}_{12z456}\!-\!P_{3z}\big[Q^{++}_{1245}\!-\!Q^{++}_{1246}\!+\!Q^{++}_{1256}\!-\!Q^{++}_{2456}\!+\!Q^{++}_{1456}\big]\big\}\nonumber\\
&&+L_{3454z}\big\{(1\!-\!P_{z4})X^{++}_{123z56}\!-\!O_{z4}X^{--}_{123z56}\!-\!P_{z4}\big[Q^{++}_{1236}\!-\!Q^{++}_{1235}\!+\!Q^{++}_{1256}\!-\!Q^{++}_{1356}\!+\!Q^{++}_{2356}\big]\big\}\nonumber\\
&&+L_{4565z}\big\{(1\!-\!P_{5z})X^{++}_{1234z6}\!-\!O_{5z}X^{--}_{1234z6}\!-\!P_{5z}\big[Q^{++}_{1234}\!-\!Q^{++}_{1246}\!+\!Q^{++}_{1236}\!-\!Q^{++}_{2346}\!+\!Q^{++}_{1346}\big]\big\}\nonumber\\
&&+L_{1656z}\big\{(1\!-\!P_{z6})X^{++}_{12345z}\!-\!O_{z6}X^{--}_{12345z}\!-\!P_{z6}\big[Q^{++}_{1234}\!-\!Q^{++}_{1235}\!+\!Q^{++}_{1245}\!-\!Q^{++}_{1345}\!+\!Q^{++}_{2345}\big]\big\}\nonumber\\
&&+L_{1623z}\big\{(P_{2z}\!-\!P_{1z})\big[Q^{++}_{z345}\!-\!Q^{++}_{z346}\!-\!Q^{++}_{z456}\!+\!Q^{++}_{z356}\big]\!-\!P_{12}Q^{++}_{3456}\!-\!O_{12}Q^{--}_{3456}\big\}\nonumber\\
&&+L_{1243z}\big\{(P_{3z}\!-\!P_{2z})\big[Q^{++}_{z456}\!-\!Q^{++}_{1z45}\!-\!Q^{++}_{1z56}\!+\!Q^{++}_{1z46}\big]\!-\!P_{23}Q^{++}_{1456}\!-\!O_{32}Q^{--}_{1456}\big\}\nonumber\\
&&+L_{2354z}\big\{(P_{4z}\!-\!P_{3z})\big[Q^{++}_{12z5}\!-\!Q^{++}_{12z6}\!-\!Q^{++}_{z265}\!+\!Q^{++}_{1z56}\big]\!-\!P_{34}Q^{++}_{1256}\!-\!O_{34}Q^{--}_{1256}\big\}\nonumber\\
&&+L_{3465z}\big\{(P_{5z}\!-\!P_{4z})\big[Q^{++}_{12z6}\!-\!Q^{++}_{13z6}\!-\!Q^{++}_{123z}\!+\!Q^{++}_{z326}\big]\!-\!P_{45}Q^{++}_{1236}\!-\!O_{54}Q^{--}_{1236}\big\}\nonumber\\
&&+L_{1645z}\big\{(P_{6z}\!-\!P_{5z})\big[Q^{++}_{123z}\!-\!Q^{++}_{124z}\!-\!Q^{++}_{z234}\!+\!Q^{++}_{1z43}\big]\!-\!P_{56}Q^{++}_{1234}\!-\!O_{56}Q^{--}_{1234}\big\}\nonumber\\
&&+L_{1265z}\big\{(P_{1z}\!-\!P_{6z})\big[Q^{++}_{z234}\!-\!Q^{++}_{z235}\!-\!Q^{++}_{z345}\!+\!Q^{++}_{z245}\big]\!-\!P_{16}Q^{++}_{2345}\!-\!O_{16}Q^{--}_{2345}\big\} ,
\eeq
where 
\beq
\bar{Q}_{1234}\bar{Q}_{5678}=Q^{++}_{1234}Q^{++}_{5678} \!+\! Q^{+-}_{1234}Q^{+-}_{5678} \!+\!Q^{--}_{1234}Q^{--}_{5678} \!+\!Q^{-+}_{1234}Q^{-+}_{5678} 
\eeq
is introduced as a shorthand notation.

We have also calculated the evolution of the octupole
\beq
\label{Sev}
&&\hspace{-0.3cm}\frac{d}{dY}\Sigma_{12345678}=\frac{\bar{\alpha_s}}{2\pi}\int_z \!-\frac{1}{2}\big(M_{12z}\!+\!M_{34z}\!+\!M_{56z}\!+\!M_{78z}\!+\!M_{18z}\!+\!M_{23z}\!+\!M_{45z}\!+\!M_{67z}\big)\Sigma_{12345678}\nonumber\\
&&\hspace{0.5cm}+L_{1823z}d_{12}X_{345678}+L_{3245z}d_{34}X_{125678}+L_{5467z}d_{56}X_{123478}+L_{1867z}d_{78}X_{123456}\nonumber\\
&&\hspace{0.5cm}+L_{1287z}d_{18}X_{723456}+L_{1243z}d_{32}X_{145678}+L_{3465z}d_{54}X_{123678}+L_{5687z}d_{76}X_{123458}\nonumber\\
&&\hspace{0.5cm}-L_{1854z}Q_{1234}Q_{5678}-L_{2367z}Q_{1278}Q_{3456}-L_{1256z}Q_{5234}Q_{1678}-L_{3478z}Q_{1238}Q_{5674}\nonumber\\
&&\hspace{0.5cm}-L_{1245z}Q_{z234}X_{1z5678}-L_{1267z}Q_{1z78}X_{z23456}-L_{3481z}Q_{123z}X_{z45678}-L_{3467z}Q_{z456}X_{123z78}\nonumber\\
&&\hspace{0.5cm}-L_{1865z}Q_{z678}X_{12345z}-L_{2356z}Q_{345z}X_{12z678}-L_{2378z}Q_{12z8}X_{34567z}-L_{4578z}Q_{567z}X_{1234z8}\nonumber\\
&&\hspace{0.5cm}+L_{1218z}d_{1z}\Sigma_{z2345678}+L_{1232z}d_{z2}\Sigma_{1z345678}+L_{2343z}d_{3z}\Sigma_{12z45678}+L_{3454z}d_{z4}\Sigma_{123z5678}\nonumber\\
&&\hspace{0.5cm}+L_{4565z}d_{5z}\Sigma_{1234z678}+L_{5676z}d_{z6}\Sigma_{12345z78}+L_{7678z}d_{7z}\Sigma_{123456z8}+L_{8781z}d_{z8}\Sigma_{1234567z}.\nonumber\\
\eeq
The evolution equation for 8-point $++$-Reggeon is a long expression. 

\beq
&&\!\!\frac{d}{dY}\Sigma^{++}_{12345678}=\frac{\bar{\alpha_s}}{2\pi}\int_z \Big[\chi^{(1)}+\chi^{(2)}+\chi^{(3)}+\chi^{(4)}+\chi^{(5)}\Big],
\eeq
where $\chi^{(1)}$ is the term that arises from the direct $\Sigma^{++}$ subtraction and comes with the same kernel as the first term in Eq.\eqref{Sev}. It is given as
\beq
\chi^{(1)}=-\frac{1}{2}\big[M_{12z}\!+\!M_{23z}\!+\!M_{34z}\!+\!M_{45z}\!+\!M_{56z}\!+\!M_{67z}\!+\!M_{78z}\!+\!M_{18z}\big]\!\Sigma^{++}_{12345678} .
\eeq
$\chi^{(2)}$ is the term that arises from dipole times hexapole terms in the evolution for $\Sigma$ whose explicit form reads 

\beq
&&\chi^{(2)}=L_{1823z} \big[( P_{2z} \!-\! P_{1z} ) ( X^{++}_{34567z} \!-\! X^{++}_{z34568} \!+\! X^{++}_{z34578} \!-\! X^{++}_{z34678} \!+\! X^{++}_{z35678} \!-\! X^{++}_{z45678} )\nonumber\\
&&\hspace{10cm}\!-\! X^{++}_{345678} P_{12} \!-\! X^{--}_{345678} O_{12} \big] \nonumber\\
&&\hspace{1cm}\!\!+L_{3245z}\big[( P_{4z}\!-\!  P_{3z} ) ( X^{++}_{12z567} \!-\! X^{++}_{12z568} \!+\! X^{++}_{12z578} \!-\! X^{++}_{12z678} \!+\! X^{++}_{1z5678} \!-\! X^{++}_{2z5678} )\nonumber\\
&&\hspace{10cm} \!-\!X^{++}_{125678} P_{34} \!-\!X^{--}_{125678} O_{34}\big]  \nonumber\\
&&\hspace{1cm}\!\!+L_{5467z}\big[( P_{6z} \!-\! P_{5z} ) ( X^{++}_{1234z7} \!-\! X^{++}_{1234z8} \!+\! X^{++}_{123z78}  \!-\! X^{++}_{124z78} \!+\! X^{++}_{134z78} \!-\! X^{++}_{234z78} )\nonumber\\
&&\hspace{10cm}\!-\! X^{++}_{123478} P_{56} \!-\! X^{--}_{123478} O_{56}  \big]\nonumber\\
&&\hspace{1cm}\!\!+ L_{1867z} \big[( P_{8z} \!-\! P_{7z} ) ( X^{++}_{12345z} \!-\! X^{++}_{12346z} \!+\! X^{++}_{12356z} \!-\! X^{++}_{12456z} \!+\! X^{++}_{13456z} \!-\! X^{++}_{z23456} )\nonumber\\
&& \hspace{10cm}\!-\! X^{++}_{123456} P_{78} \!-\!  X^{--}_{123456} O_{78} \big]\nonumber\\
&&\hspace{1cm}\!\!+L_{1287z} \big[ ( P_{1z} \!-\! P_{8z} ) ( X^{++}_{23456z} \!-\! X^{++}_{23457z} \!+\! X^{++}_{23467z} \!-\! X^{++}_{23567z} \!+\! X^{++}_{24567z} \!-\! X^{++}_{34567z} ) \nonumber\\
&& \hspace{10cm}\!-\!X^{++}_{234567} P_{18} \!-\! X^{--}_{234567} O_{18}  \big] \nonumber\\
&&\hspace{1cm}\!\!+L_{1243z} \big[( P_{2z} \!-\! P_{3z} ) ( X^{++}_{1z4567} \!-\! X^{++}_{1z4568} \!+\! X^{++}_{1z4578} \!-\! X^{++}_{1z4678} \!+\! X^{++}_{1z5678} \!-\! X^{++}_{z45678} )\nonumber\\
&& \hspace{10cm}\!-\!X^{++}_{145678} P_{32} \!-\! X^{--}_{145678} O_{32} \big] \nonumber\\
&&\hspace{1cm}\!\!+L_{3465z}\big[( P_{4z} \!-\! P_{5z} ) ( X^{++}_{123z67} \!-\! X^{++}_{123z68} \!+\! X^{++}_{123z78} \!-\! X^{++}_{12z678} \!+\! X^{++}_{13z678} \!-\! X^{++}_{23z678} )\nonumber\\
&& \hspace{10cm} \!-\! X^{++}_{123678} P_{54} \!-\!  X^{--}_{123678} O_{54} \big] \nonumber\\
&&\hspace{1cm}\!\!+L_{5687z} \big[( P_{6z} \!-\! P_{7z} ) ( X^{++}_{12345z} \!-\! X^{++}_{1234z8} \!+\! X^{++}_{1235z8} \!-\! X^{++}_{1245z8} \!+\! X^{++}_{1345z8} \!-\! X^{++}_{2345z8} )\nonumber\\
&&  \hspace{10cm}  \!-\!X^{++}_{123458}P_{76} \!-\! X^{--}_{123458}O_{76}  \big] . \nonumber\\
\eeq
The $\chi^{(3)}$ part of the evolution of 8-point $++$-Reggeon originates from the quadrupole square terms in the evolution of $\Sigma$. Its explicit form is

\beq
&&\chi^{(3)}=L_{1854z} \big[ \bar{Q}_{1234}\bar{Q}_{5678}  \!+\!( Q^{++}_{123z} \!-\! Q^{++}_{124z} \!+\! Q^{++}_{1z43} \!-\! Q^{++}_{z234} ) ( Q^{++}_{z567} \!-\! Q^{++}_{z568} \!+\! Q^{++}_{z578} \!-\! Q^{++}_{z678} ) \big]\nonumber\\ 
&&\hspace{1cm}\!\!+L_{2367z} \big[ \bar{Q}_{1278}\bar{Q}_{3456}  \!+\! ( Q^{++}_{z345} \!-\! Q^{++}_{z346} \!+\! Q^{++}_{z356} \!-\! Q^{++}_{z456} ) ( Q^{++}_{1z78} \!-\! Q^{++}_{z287} \!+\! Q^{++}_{12z7} \!-\! Q^{++}_{12z8} ) \big] \nonumber \\ 
&&\hspace{1cm}\!\!+L_{1256z} \big[ \bar{Q}_{1678}\bar{Q}_{2345} \!+\! ( Q^{++}_{z234} \!-\! Q^{++}_{z235} \!+\! Q^{++}_{z245} \!-\! Q^{++}_{z345} ) ( Q^{++}_{z678} \!-\! Q^{++}_{1z67} \!+\! Q^{++}_{1z68} \!-\! Q^{++}_{1z78} ) \big] \nonumber\\ 
&&\hspace{1cm}\!\!+L_{3478z} \big[ \bar{Q}_{1238}\bar{Q}_{4567}  \!+\!( Q^{++}_{z456} \!-\! Q^{++}_{z457} \!+\! Q^{++}_{z467} \!-\! Q^{++}_{z567} ) ( Q^{++}_{z328} \!-\! Q^{++}_{13z8} \!+\! Q^{++}_{12z8} \!-\! Q^{++}_{123z} ) \big] . \nonumber \\  
\eeq
Here we use again the shorthand notation 
\beq
\bar{Q}_{1234}\bar{Q}_{5678}=Q^{++}_{1234}Q^{++}_{5678} \!+\! Q^{+-}_{1234}Q^{+-}_{5678} \!+\!Q^{--}_{1234}Q^{--}_{5678} \!+\!Q^{-+}_{1234}Q^{-+}_{5678} .
\eeq
$\chi^{(4)}$ originates from quadrupole times hexapole terms in the evolution of the 8-point $++$-Reggeon. 
\beq
&&\chi^{(4)}=L_{1245z} \big[ \bar{X}_{1z5678} \bar{Q}_{z234} \!+\!X^{++}_{1z5678} ( P_{23} \!-\! P_{24} \!+\! P_{34} )\nonumber\\
&&\hspace{6cm} \!+\! (  Q^{++}_{1568} \!-\! Q^{++}_{1567} \!+\! Q^{++}_{1678} \!-\! Q^{++}_{1578}  \!+\! Q^{++}_{5678} ) Q^{++}_{z234} \big]\nonumber\\
&&\hspace{1cm}\!\!+L_{1267z} \big[ \bar{X}_{23456z}\bar{Q}_{1z78} \!+\!  X^{++}_{23456z} ( P_{18} \!-\! P_{17} \!+\! P_{78} ) \nonumber\\
&& \hspace{6cm} \!+\!  ( Q^{++}_{2345} \!-\! Q^{++}_{2346} \!+\! Q^{++}_{2356} \!-\! Q^{++}_{2456} \!+\! Q^{++}_{3456} ) Q^{++}_{1z78} \big]\nonumber\\
&&\hspace{1cm}\!\!+L_{3481z}\big[ \bar{X}_{z45678}\bar{Q}_{123z} \!+\! X^{++}_{z45678} ( P_{12} \!-\! P_{13} \!+\! P_{23} )\nonumber\\
&& \hspace{6cm} \!+\! ( Q^{++}_{4567} \!-\! Q^{++}_{4568} \!+\! Q^{++}_{4578} \!-\! Q^{++}_{4678} \!+\! Q^{++}_{5678} ) Q^{++}_{123z} \big] \nonumber\\
&&\hspace{1cm}\!\!+L_{3467z} \big[ \bar{X}_{123z78} \bar{Q}_{z456} \!+\! X^{++}_{123z78} (P_{45} \!-\! P_{46} \!+\! P_{56} ) \nonumber\\
&& \hspace{6cm}  \!+\! ( Q^{++}_{1238} \!-\! Q^{++}_{1237}  \!+\! Q^{++}_{1278} \!-\! Q^{++}_{1378} \!+\! Q^{++}_{2378} ) Q^{++}_{z456} \big]\nonumber\\ 
&&\hspace{1cm}\!\!+L_{1865z} \big[ \bar{X}_{12345z}\bar{Q}_{z678} \!+\! X^{++}_{12345z} ( P_{67} \!-\! P_{68} \!+\! P_{78} ) \nonumber\\
&& \hspace{6cm}    \!+\! ( Q^{++}_{1234} \!-\! Q^{++}_{1235} \!+\! Q^{++}_{1245} \!-\! Q^{++}_{1345} \!+\! Q^{++}_{2345} ) Q^{++}_{z678} \big]\nonumber\\
&& \hspace{1cm} \!\!+L_{2356z} \big[ \bar{X}_{12z678} \bar{Q}_{z345} \!+\! X^{++}_{12z678} (P_{34} \!-\! P_{35} \!+\! P_{45} )  \nonumber\\
&& \hspace{6cm}   \!+\! ( Q^{++}_{1267} \!-\! Q^{++}_{1268} \!+\! Q^{++}_{1278} \!-\! Q^{++}_{2678}  \!+\! Q^{++}_{1678}) Q^{++}_{z345} \big] \nonumber\\
&& \hspace{1cm} \!\!+L_{2378z} \big[ \bar{X}_{34567z} \bar{Q}_{12z8} \!+\! X^{++}_{34567z}  (P_{12} \!-\! P_{28} \!+\! P_{18} )  \nonumber\\
&& \hspace{6cm}   \!+\! ( Q^{++}_{3456} \!-\! Q^{++}_{3457} \!+\! Q^{++}_{3467} \!-\! Q^{++}_{3567} \!+\! Q^{++}_{4567} ) Q^{++}_{12z8} \big] \nonumber\\
&&  \hspace{1cm} \!\!+L_{4578z} \big[ \bar{X}_{1234z8}\bar{Q}_{z567} \!+\! X^{++}_{1234z8} ( P_{56} \!-\! P_{57} \!+\! P_{67} ) \nonumber\\
&& \hspace{6cm}   \!+\! ( Q^{++}_{1234} \!+\! Q^{++}_{1238} \!-\! Q^{++}_{1248} \!+\! Q^{++}_{1348} \!-\! Q^{++}_{2348} ) Q^{++}_{z567} \big],\nonumber\\
\eeq
where we introduced the following shorthand notation
\beq
\bar{X}_{12345z} \bar{Q}_{z678}=X^{++}_{12345z} Q^{++}_{z678}+X{+-}_{12345z} Q^{+-}_{z678}+X^{--}_{12345z} Q^{--}_{z678}+X^{-+}_{12345z} Q^{-+}_{z678}.
\eeq

Finally, $\chi^{(5)}$ are the terms that come from the dipole times hexapole terms in the evolution of the 8-point function. 
Its explicit form reads

\beq
&&\chi^{(5)}=L_{1218z} \big[ P_{1z} ( \!-\! X^{++}_{234567} \!+\! X^{++}_{234568} \!-\! X^{++}_{234578} \!+\! X^{++}_{234678} \!-\! X^{++}_{235678} \!+\! X^{++}_{245678} \!-\! X^{++}_{345678} )\nonumber\\
&&\hspace{8.8cm} +( 1 \!-\! P_{1z})\Sigma^{++}_{z2345678}  \!-\! O_{1z}\Sigma^{--}_{z2345678} \big]\nonumber\\
&&\hspace{1cm}\!\!+L_{1232z} \big[ P_{2z} (  X^{++}_{134567} \!-\! X^{++}_{134568} \!+\! X^{++}_{134578} \!-\! X^{++}_{134678} \!+\! X^{++}_{135678} \!-\! X^{++}_{145678} \!-\! X^{++}_{345678} ) \nonumber\\
&&\hspace{8.8cm}   + (1 \!-\!  P_{2z}) \Sigma^{++}_{1z345678} \!-\! O_{z2}\Sigma^{--}_{1z345678}\big]\nonumber\\
&&\hspace{1cm}\!\!+L_{2343z}\big[ P_{3z} ( \!-\! X^{++}_{124567} \!+\! X^{++}_{124568} \!-\! X^{++}_{124578} \!+\! X^{++}_{124678} \!-\! X^{++}_{125678} \!-\! X^{++}_{145678} \!+\! X^{++}_{245678} ) \nonumber\\
&&\hspace{8.8cm} +( 1\!-\!  P_{3z}) \Sigma^{++}_{12z45678} \!-\! O_{3z}\Sigma^{--}_{12z45678}\big]\nonumber\\
&&\hspace{1cm}\!\!+L_{3454z} \big[ P_{4z} ( X^{++}_{123567} \!-\! X^{++}_{123568} \!+\! X^{++}_{123578} \!-\! X^{++}_{123678} \!-\! X^{++}_{125678} \!+\! X^{++}_{135678} \!-\! X^{++}_{235678} )  \nonumber\\
&&\hspace{8.8cm}  +( 1 \!-\! P_{4z})\Sigma^{++}_{123z5678}  \!-\! O_{z4}\Sigma^{--}_{123z5678} \big]\nonumber\\
&&\hspace{1cm}\!\!+L_{4565z} \big[ P_{5z} ( \!-\!X^{++}_{123467} \!+\! X^{++}_{123468} \!-\! X^{++}_{123478} \!-\! X^{++}_{123678} \!+\! X^{++}_{124678} \!-\! X^{++}_{134678} \!+\! X^{++}_{234678} ) \nonumber\\
&&\hspace{8.8cm} + \Sigma^{++}_{1234z678} ( 1 \!-\! P_{5z})  \!-\! O_{5z}\Sigma^{--}_{1234z678} \big] \nonumber\\
&&\hspace{1cm}\!\!+L_{5676z} \big[ P_{6z} ( X^{++}_{123457} \!-\! X^{++}_{123458} \!-\! X^{++}_{123478} \!+\! X^{++}_{123578} \!-\! X^{++}_{124578} \!+\! X^{++}_{134578} \!-\! X^{++}_{234578} ) \nonumber\\
&&\hspace{9cm}    \!+\! ( 1 \!-\! P_{z6}) \Sigma^{++}_{12345z78}   \!-\!  O_{z6}\Sigma^{--}_{12345z78} \big] \nonumber\\
&&\hspace{1cm}\!\!+L_{6787z} \big[ P_{7z} ( \!-\! X^{++}_{123456} \!-\! X^{++}_{123458} \!+\! X^{++}_{123468} \!-\! X^{++}_{123568} \!+\! X^{++}_{124568} \!-\! X^{++}_{134568} \!+\!X^{++}_{234568} )  \nonumber\\
&&\hspace{8.8cm}  + ( 1\!-\!  P_{7z})  \Sigma^{++}_{123456z8}\!-\! O_{7z}\Sigma^{--}_{123456z8}\big] \nonumber\\
&&\hspace{1cm} \!\!+L_{1878z} \big[ P_{8z} ( \!-\! X^{++}_{123456} \!+\! X^{++}_{123457} \!-\! X^{++}_{123467} \!+\! X^{++}_{123567} \!-\! X^{++}_{124567} \!+\! X^{++}_{134567} \!-\! X^{++}_{234567} )  \nonumber\\
&&\hspace{8.8cm}  + ( 1\!-\! P_{8z} ) \Sigma^{++}_{1234567z}  \!-\! O_{z8}\Sigma^{--}_{1234567z}\big] . \nonumber\\
\eeq

\section*{Acknowledgments}
T.A. and M.L. thank the Physics Department of the University of Connecticut for hospitality during the visits while the work 
on this project was in progress.  A.K. and T.A. thank  the  Physics Departments of the Ben-Gurion University of the Negev; A.K. also thanks 
  Universidad T\'ecnica Federico Santa Mar\'\i a.
The research  was supported by the DOE grant DE-FG02-13ER41989; the BSF grant 2012124,    Marie Curie Grant  PIRG-GA-2009-256313; the  ISRAELI SCIENCE FOUNDATION grant \#87277111;  the People Programme (Marie Curie Actions) of the European Union's Seventh Framework Programe FP7/2007-2013/ under REA
grant agreement \%318921;  the  Fondecyt (Chile) grants  1100648 and 1130549; European Research Council grant HotLHC ERC-2001-
StG-279579; Ministerio de Ciencia e Innovac\'\i on of Spain grants  FPA2011-22776 and Consolider-Ingenio 2010 CPAN CSD2007-00042 and by FEDER.

\appendix

\bibliographystyle{jhep}

\end{document}